\documentclass[aps,showpacs,amsmath,amssymb]{revtex4}
\newtheorem{theorem}{Theorem}
\newtheorem{lemma}[theorem]{Lemma}
\newtheorem{definition}[theorem]{Definition}

\newtheorem{remark}[theorem]{Remark}
\newtheorem{cor}[theorem]{Corollary}
\newtheorem{pro}[theorem]{Proposition}
\usepackage[dvips]{graphicx}
\usepackage{dcolumn}
\usepackage[english]{babel}
\usepackage{bm}
\usepackage{amsmath}
\usepackage{amssymb}
\usepackage[all]{xy}
\begin{document}
\title[Spectrums]{Spectral Relationships Between Kicked Harper \\ and On--Resonance Double Kicked Rotor Operators}
\author{Wayne Lawton}
\email{wlawton@math.nus.edu.sg}
\affiliation{Department of Mathematics,
National University of Singapore, 2~Science~Drive~2, Singapore 117543}
\author{Anders S. Mouritzen}
\email{asm@phys.au.dk}
\affiliation{Department of Physics and Center of Computational Science and Engineering,
National University of Singapore, 117542, Singapore}
\affiliation{Department of Physics and Astronomy,
University of Aarhus, DK-8000, Aarhus C, Denmark}
\author{Jiao Wang}
\affiliation{Temasek Laboratories, National University of Singapore, 117542, Singapore}
\affiliation{Beijing-Hong Kong-Singapore Joint Center for
Nonlinear and Complex Systems (Singapore),
National University of Singapore, 117542, Singapore}
\author{Jiangbin Gong}
\affiliation{Department of Physics and Center of Computational Science and Engineering,
National University of Singapore, 117542, Singapore}
\affiliation{NUS Graduate School for Integrative Sciences
and Engineering, Singapore 117597, Singapore}

\date{\today}
%
%
\def\BB{{\bf B}}
\def\CC{{\bf C}}
\def\DD{{\bf D}}
\def\GG{{\bf G}}
\def\HH{{\bf H}}
\def\NN{{\bf N}}
\def\QQ{{\bf Q}}
\def\RR{{\bf R}}
\def\TT{{\bf T}}
\def\ZZ{{\bf Z}}
\def\cA{{\cal A}}
\def\cB{{\cal B}}
\def\cC{{\cal C}}
\def\cF{{\cal F}}
\def\cH{{\cal H}}
\def\cK{{\cal K}}
\def\cM{{\cal M}}
\def\cZ{{\cal Z}}
\begin{abstract}
Kicked Harper operators and on--resonance
double kicked rotor operators model quantum systems whose semiclassical limits
exhibit chaotic dynamics. Recent computational studies indicate a striking resemblance between the
spectrums of these operators. In this paper we apply $C^*$--algebra
methods to explain this resemblance. We show that each pair of
corresponding operators belong to a common rotation $C^*$--algebra
$\cB_{\alpha}$, prove that their spectrums are equal if $\alpha$ is
irrational, and prove that the Hausdorff distance between their
spectrums converges to zero as $q$ increases if $\alpha =
p/q$ with $p$ and $q$ coprime integers. Moreover,
we show that corresponding operators in $\cB_\alpha$ are homomorphic images of
mother operators in the universal rotation $C^*$--algebra $\cA_\alpha$
that are unitarily equivalent and hence have identical spectrums. These results
extend analogous results for almost Mathieu operators. We also utilize the $C^*$--algebraic
framework to develop efficient algorithms to compute the spectrums of these mother operators
for rational $\alpha$ and present preliminary numerical results that support the conjecture that their spectrums are Cantor sets if $\alpha$ is irrational. This conjecture for almost Mathieu operators,
called the Ten Martini Problem, was recently proved after intensive
efforts over several decades. This proof for the almost Mathieu operators utilized transfer matrix methods, which do not exist for the kicked operators. We outline a strategy, based on a
special property of loop groups of semisimple Lie groups, to prove
this conjecture for the kicked operators.
\end{abstract}


\maketitle
\newpage
\section{Introduction}
\label{intro}
The families of kicked Harper and on--resonance double kicked rotor operators that we discuss in this paper arose comparatively recently in the field of quantum chaos and most of the knowledge about them has been acquired through numerical computation. They are related to another family of operators that arose earlier, that is simpler, and whose properties are far better understood theoretically. We start by reviewing this family of operators.
\\ \\
We let $\CC,$ $\RR,$ $\ZZ,$ and $\NN = \{1,2,3,4,...\}$ denote the complex, real, integer, and natural numbers. Furthermore, we let $\TT = \RR\, / \, \ZZ$ denote the circle group parameterized by the interval $[0,1),$ and $\TT_{c}$ denote the circle group parameterized by the numbers on the complex unit circle. The family of almost Mathieu self-adjoint operators are represented on $L^2(\TT).$ The representation with respect to the standard orthonormal basis $\{ \, \xi_n(x) = \exp (i2 \pi n x) \ : \ n \in \ZZ \, \}$ is
\begin{equation}\label{eq:H1}
    H(\alpha,\lambda,\theta) \, \xi_n =
        \xi_{n+1} + \xi_{n-1} + 2 \, \lambda \, \cos \big[ 2 \pi (n \alpha + \theta) \big] \ \xi_{n}, \ \ n \in \ZZ.
\end{equation}
Here the coupling constant $\lambda \in \RR \backslash \{0\},$ the frequency $\alpha \in [0,1),$ and the phase $\theta \in [0,1).$
These (also called Harper) operators were introduced in 1955 by Harper \cite{harper55} to explain the effect of magnetic fields on conduction bands of a metal. The spectrums of these operators have been studied for several decades and are well understood.
In 1964 Azbel \cite{azbel64} conjectured that if $\alpha$ is irrational then the spectrum of $H(\alpha,\lambda,\theta)$ is a Cantor set.
In 1976 Hofstadter \cite{hofstadter76}, observing that the spectrum of $H(\alpha,\lambda,\theta)$ is
independent of $\theta$ if and only if $\alpha$ is irrational, argued that "there can be no physical effect stemming from the irrationality of some parameter" and proposed to study instead the set
\begin{equation}
\label{barsigma}
    S(\alpha,\lambda) =
        \bigcup_{\theta \in [0,1)} \, \textnormal{spectrum}\big[H(\alpha,\lambda,\theta)\big].
\end{equation}
He computed $S(\alpha,1)$ numerically for a fine
grid of rational values of $\alpha \in (0,1)$ and studied the graph
of this set-valued function of $\alpha$ (a subset of the rectangle
$[-4,4] \times (0,1)$). He asserted that the graph had a
{\it butterfly pattern}, discussed the
relationship between the recursive clustering structure of the graph
and the continued fraction expansion of $\alpha.$
In 1981 the problem of proving Azbel's conjecture was named the Ten Martini Problem by Barry Simon after an offer by Mark Kac \cite{simon83}.
In 1982 Simon and Bellisard \cite{bellissardsimon82} showed that if $\alpha = p/q$ where $p$ and $q$ are coprime, then $S(\alpha,\lambda)$ consists of $q$ disjoint closed intervals if $q$ is odd and $q-1$ disjoint closed intervals if $q$ is even, and they used this result to prove that for a $G_\delta$ set of pairs $(\alpha,\lambda)$ in $\RR^2,$ $S(\alpha,\lambda)$ is a Cantor set.
In 1993 Choi, Elliot and Yui \cite{choielliotyui90} used the noncommutative binomial theorem to compute sharper estimates for the spectral gaps when $\alpha$ is rational and used these estimates to prove that $S(\alpha,\lambda)$ is a Cantor set whenever $\alpha$ is a Liouville number.
In 1993 Last \cite{last94} proved that for almost all $\alpha \in (0,1)$ the Lebesque measure of $S(\alpha,\lambda)$ equals $4\big|1 - |\lambda|\, \big|$ for all $\lambda \in \RR,$ and hence that $S(\alpha,1)$ is a Cantor set.
In 1987 Sinai \cite{sinai87} used KAM theory to prove that $S(\alpha,\lambda)$ is a Cantor set for almost all $\alpha$ and sufficiently small (or large) $|\lambda|.$
The Ten Martini Problem was recently proved by Avila and Jitomirskaya \cite{avilajitomirskaya08}.
Closely related to the Harper operators, we define the family of unitary Harper operators
\begin{equation}\label{eq:Uh1}
    U_{H}(\kappa,\alpha,\lambda,\theta) =
        \exp \big[ -i \, \kappa \, H(\alpha,\lambda,\theta) \, \big]
\end{equation}
where $\kappa \in \RR$ and $\alpha, \lambda, \theta$ are as in Equation~\ref{eq:H1}. We will show in Corollary~\ref{UHcantor} that if $\alpha$ is irrational, then the spectrum of $U_{H}$ is independent of $\theta$ and is a Cantor set.
\\ \\
In 1979 Casati, Chirikov, Ford and Izrailev \cite{casati79} and
Berry, Balazs, Taber and Voros \cite{berry79} initiated the study of
families of unitary operators that describe the time evolution of
quantum systems whose classical limits exhibit chaotic dynamics.
Developments in the field of quantum chaos \cite{casatibook95} led
to the so-called kicked Harper operators that admit the following
representations on $L^2(\TT):$
\begin{equation}\label{eq:Ukh1}
    U_{kH}(\kappa,\alpha,\lambda,\theta) =
        \exp \Big[ -i 2 \kappa \, \cos(2\pi x) \, \Big] \,
        \exp \Big[ -i 2 \kappa \, \lambda \cos \Big(-i \alpha \frac{d}{dx} + 2\pi \theta \Big) \Big]. \,
\end{equation}
We observe that $U_{kH}(\kappa,\alpha,\lambda,\theta)
\rightarrow U_H(\kappa,\alpha,\lambda,\theta)$ as $|\kappa|
\rightarrow 0.$
%
%
%
%
In 1990 Leboeuf, Kurchan, Feingold and Arovas
\cite{leboeuf90} studied a quantum version of a classical kicked
Harper map whose phase space is a (compact) torus and that exhibits
chaotic dynamics for sufficiently large values of certain
parameters. Their quantum kicked Harper operators arise through
canonical quantization of this classical Harper map and have the
form
$\widetilde {U}_{kH}(\kappa,\alpha,\lambda),$
(see Equation~\ref{op4} below) and is represented on the Hilbert
space $L^2(\TT^2).$ To quantize the system on the torus they set
$\alpha$ to be the area of classical phase space in units of
Planck's constant and they choose $\alpha = 1/q$ with $q$ an
integer. However, their quantum operators admit the family
(parameterized by $\theta$) of representations on $L^2(\TT)$ that
have the form $U_{kH}(\kappa,1/q,\lambda,\theta).$
In 1991 Geisel, Ketzmerick and Petschel \cite{geisel91} studied the
dependence on $\kappa$ of the spectra of the operators
$\widetilde {U}_{kH}(\kappa,\alpha,1),$
where $\alpha$ ranges over the interval $[0,1)$. These spectra are
the same as the union over $\theta$ of the spectrums of $U_{kH}(\kappa,\alpha,\lambda,\theta).$ They also remark that $\theta$ can be thought of as being proportional to a quasimomentum which arises in analogy to Bloch's theorem.
They observe that for small $|\kappa|$ the collection of spectrums for rational $\alpha \in [0,1)$ has the Hofstadter butterfly pattern for small $|\kappa|$, and that as $|\kappa|$ increases the spectral bands close up. They also compute the effect of $\kappa$ on other quantities such as diffusion and localization.
In 1992 Artuso, Casati, and Shepelyansky \cite{artuso92} studied a quantum version
of another classical kicked Harper map whose
phase space is a (noncompact) cylinder. Their operators have the form
$U_{kH}(\kappa,\alpha,\lambda,0).$
This quantization evidently differs from the canonical quantization
of Leboeuf {\it et al.} insofar as the state space of the quantum
system equals $L^2(\TT)$, the quasimomentum parameter is fixed at
$\theta=0$, and there are no restrictions on
$\alpha$. The quantum version with $\theta=0$ was later widely used,
though a more general quantization gives the operator $U_{kH}(\kappa,\alpha,\lambda,\theta)$ \cite{Borgonovi95}.
The fact that a classical kicked Harper map can be quantized in many ways was
probably best summarized by Guarneri and Borgonovi
\cite{Guarneri93}.

Another family of operators we study here is the so-called
on--resonance double kicked rotor operators.  They admit the
following representations on $L^2(\TT)$:
\begin{equation}\label{eq:Uordkr1}
    U_{ordkr}(\kappa,\alpha,\lambda,\theta) =
        \exp \Big[ -i 2 \kappa \, \cos(2\pi x) \, \Big] \,
        T(\alpha) \, \exp \Big[ -i 2 \kappa \, \lambda \, \cos\big(2\pi(x+\theta)\big) \, \Big] \, T(\alpha)^{-1},
\end{equation}
where
$T(\alpha) = \exp \left( \frac{-i\alpha}{4\pi} \frac{d^2}{dx^2}
\right).$
The operators $U_{ordkr}(\kappa,\alpha,\lambda,0)$ were introduced
by Gong and Wang in \cite{gongwang07}, where they compared their
properties with those of operators
$U_{kH}(\kappa,\alpha,\lambda,0).$ In \cite{wang08} they discussed
how to implement the operators $U_{ordkr}(\kappa,\alpha,\lambda,\theta)$
for $\theta \neq 0$ and related $\theta$ to a symmetry breaking
condition in a Hamiltonian ratchet model. In \cite{wanggong08} and
\cite{wangmouritzengong08} they and Mouritzen discussed how the
operators $U_{ordkr}(\kappa,\alpha,\lambda,0)$ can be realized
experimentally,
and demonstrated numerically a striking resemblance between the
spectrums of $U_{kH}(1/2,\alpha,1,0)$ and
$U_{ordkr}(1/2,\alpha,1,0)$ for a large set of rational
$\alpha = p/q$ with $p$ and $q$ coprime. In particular,
Figure~2 in \cite{wanggong08} shows that the spectrums for $\alpha =
13/41$ are different, but very nearly equal.

One main objective of this work is to mathematically explain this
resemblance between on-resonance double kicked rotor operators and
the kicked Harper operators, by proving that for all values of
$(\kappa, \lambda, \theta_1, \theta_2)$ the spectrums of
$U_{ordkr}(\kappa,\alpha,\lambda,\theta_1)$ and
$U_{kH}(\kappa,\alpha,\lambda,\theta_2)$ are equal whenever $\alpha$
is irrational and that their Hausdorff distance approaches zero as
$q$ increases for rational $\alpha = p/q.$ Because Wang
and Gong \cite{wanggong08} also showed that the dynamics of the
on-resonance double kicked rotor model dramatically differs from
that of the kicked Harper model for irrational $\alpha$, our proof
of the spectral equivalence, in addition to mathematical interest,
shall motivate new theoretical problems, such as the
characterization of (generalized) eigenfunctions of the operators and of density of states or of the degeneracy of the spectrums, which ultimately accounts for the dynamical differences
observed by Wang and Gong.

Our proof is based on $C^{*}$--algebra methods. The proof itself
also leads to efficient algorithms for numerical studies of
the above-mentioned unitary operators. Furthermore, we discuss the
observation that the spectral graph of $U_{ordkr}(\kappa,\alpha,\lambda)$ has a butterfly
pattern similar to that previously observed for $U_{kH}(\kappa,\alpha,\lambda)$. Though
rigorous spectral characterization of $U_{kH}(\kappa,\alpha,\lambda)$ and $U_{ordkr}(\kappa,\alpha,\lambda)$
remains an open problem, we outline a strategy to prove the conjecture that
the spectrum of their mother operators $\widetilde U_{kH}(\kappa,\alpha,\lambda)$, or identically
$\widetilde U_{ordkr}(\kappa, \alpha, \lambda)$, is indeed a Cantor set for
irrational $\alpha$. These mother operators, denoted with a tilde, live in the universal rotation $C^*$--algebra $\cA_\alpha$ and the corresponding operators $U_{kH}(\kappa,\alpha,\lambda)$ and $U_{ordkr}(\kappa,\alpha,\lambda)$, living in the rotation algebra $\cB_\alpha$, are homomorphic images of
them.

Our paper is arranged as follows:
In section~\ref{derivations} we derive results concerning the spectrums of $H,$ $U_H,$ $U_{kH}$ and $U_{ordkr}$, in particular the main result that for irrational $\alpha,$ the spectrums of $U_{ordkr}(\kappa,\alpha,\lambda,\theta)$ and $U_{kH}(\kappa,\alpha,\lambda,\theta)$ are identical. In section~\ref{algorithms} we present efficient algorithms for numerically calculating the spectra of $U_{ordkr}(\kappa,\alpha,\lambda,\theta)$ and $U_{kH}(\kappa,\alpha,\lambda,\theta)$ for rational $\alpha = p/q.$ The treatment in both of these sections~\ref{derivations} and \ref{algorithms} rely on results given in appendices~\ref{appendix2} and \ref{appendix3}. In section~\ref{sec:futureresearch}, we discusses ideas for future research. Appendix~\ref{appendix1} briefly reviews the considerations and difficulties involved in physically realizing the operators above. Appendix~\ref{appendix2} summarizes results in spectral theory and the theory of $C^*$--algebras used in the paper. Appendix~\ref{appendix3} further builds on these results and summarizes results regarding rotation $C^*$--algebras used in the paper.

\section{Derivations}
\label{derivations}
For $r \geq 0$ we define the disc
$\DD(r) = \{ \, z \in \CC \, : \, |z| \leq r \, \}.$
For $S \subseteq \CC$ we let $\cH(S)$ denote the metric space whose points are
compact subsets of $S$ and that is equipped with the {\it Hausdorff metric}
\begin{equation}
\label{HD}
\textnormal{d}(X, Y) =
\max \, \big\{ \
    \max_{x \in X} \min_{y \in Y} \, |x-y| \, , \
    \max_{y \in Y} \min_{x \in X} \, |x-y| \,  \
    \big\}\, , \ \ \ \ X, Y \in \cH(S).
\end{equation}
This metric space is called the {\it Hyperspace} of $S.$ A standard result in topology ensures that the
hyperspace $\cH(S)$ is compact whenever $S$ is compact (\cite{dugundji66}, p.205, p.253), (\cite{munkres75}, p.279).
For every Hilbert space $\HH$ we denote the set of bounded operators on $\HH$ by $\cB(\HH),$
its subset of operators that are normal by $\cB_{n}(\HH),$ unitary by $\cB_{u}(\HH),$ and
self-adjoint by $\cB_{a}(\HH).$ We observe that the real part of $A \in \cB(\HH),$ given by
$\Re(A) = \frac{1}{2}(A+A^*),$ is self-adjoint. If $A \in \cB_a(\HH)$ then
$\exp (iA) \in \cB_u(\HH)$ and Lemma~$3.3$ in \cite{boca01} implies that
\begin{equation}
\label{eq:expcont}
    ||\exp (iA) - \exp(iB)|| \leq ||A-B||, \ \ A, B \in \cB_a(\HH).
\end{equation}
We define
$\sigma \, : \, \cB(\HH) \rightarrow \cH(\CC)$
by
\begin{equation}
\label{sigma}
    \sigma(B) \ = \ \textnormal{spectrum}(B) \ = \ \big\{ \, \mu \in \CC \, : \, \mu \, I - B \ \textnormal{does not have an inverse in} \ \cB(\HH) \, \big\}, \qquad B \in \cB(\HH).
\end{equation}
As noted in the Examples following Definition \ref{Cstaralgebra}, $\cB(\HH)$ is a $C^*$--algebra,
and Lemma \ref{gelfand2}, as noted in the comment following it, implies that for every
$B \in \cB(\HH),$
$\sigma(B) = \sigma_{\cB(\HH)}(B).$
Therefore, Lemma~\ref{spec2} implies that
$\sigma(B) \in \cH\big(\DD(||B||)\big)$
for every $B \in \cB(\HH).$ Furthermore,
$\sigma\big(\cB_a(\HH)\big) \subseteq \cH(\RR)$
and
$\sigma\big(\cB_u(\HH)\big) \subseteq \cH(\TT_c).$
If $\dim(\HH) < \infty$ and $A \in \cB(\HH)$ then $A$ can be represented by a matrix and $\sigma(A)$ is the set of eigenvalues of this matrix. This is not the case when $\HH$ is infinite dimensional. For example,
if $\HH = L^2([0,1])$ and $(Af)(x) = x f(x)$ then $\sigma(A) = [0,1]$ and $A$ has no eigenvalues since it has no eigenvectors. Although $\sigma$ is not continuous (\cite{kato66}, Example 3.8), Proposition~\ref{dist} shows that the restriction of $\sigma$ to $\cB_n(\HH)$ is continuous since
\begin{equation}
\label{eq:speccont}
    d\left[\sigma(A),\sigma(B)\right] \ \leq \ || \, A - B \, ||, \qquad A, B \in \cB_n(\HH).
\end{equation}
We observe that since the operators $H,$ $U_{kH},$ and $U_{ordkr}$ are normal, Equations (\ref{eq:expcont}) and (\ref{eq:speccont}) imply that their spectrums depend continuously on the parameters $\kappa, \lambda, \theta.$ In particular
\begin{eqnarray}
\label{eq:Hcont}
    d\Big[\sigma \big(H(\alpha,\lambda,\theta_1)\big), \, \sigma \big(H(\alpha,\lambda,\theta_2) \big)\Big] \
    &\leq& \ 2|\lambda| \ \big|\sin[\pi(\theta_1-\theta_2)] \big|\, , \\
\label{eq:UHcont}
    d\Big[\sigma \big(U_H(\kappa,\alpha,\lambda,\theta_1)\big),  \, \sigma \big(U_H(\kappa,\alpha,\lambda,\theta_2) \big)\Big] \
    &\leq& \ 2|\kappa \, \lambda| \ \big|\sin[\pi(\theta_1-\theta_2)] \big|\, , \\
\label{eq:UkHcont}
    d\Big[\sigma \big(U_{kH}(\kappa,\alpha,\lambda,\theta_1)\big), \, \sigma \big(U_{kH}(\kappa,\alpha,\lambda,\theta_2) \big)\Big] \
    &\leq& \ 2|\kappa \, \lambda| \ \big|\sin[\pi(\theta_1-\theta_2)]\big|\, ,\\
\label{eq:Uordkrcont}
 d\Big[\sigma \big(U_{ordkr}(\kappa,\alpha,\lambda,\theta_1)\big), \, \sigma \big(U_{ordkr}(\kappa,\alpha,\lambda,\theta_2) \big)\Big] \
    &\leq& \ 2|\kappa \, \lambda| \ \big|\sin[\pi(\theta_1-\theta_2)]\big|\, .
\end{eqnarray}
\begin{remark}
\label{remark1}
We observe that if
$\alpha_1 \not \in \ZZ,$ $\alpha_2 \not \in \ZZ,$
$\alpha_1 - \alpha_2 \not \in \ZZ,$ and $\alpha_1 + \alpha_2 \not \in \ZZ,$
then
\begin{equation}
\label{diff}
    || \, H(\alpha_1,\lambda,\theta) - H(\alpha_2,\lambda,\theta) \, || \ = \
            \sup_{n \in \ZZ} \, \Big| \, 2\lambda \,
                \sin \big[\pi n (\alpha_1 + \alpha_2) + 2\pi \theta \big]\, \sin \big[\pi n (\alpha_1 - \alpha_2)\big]\, \Big|
                    \ \geq \ \hbox{$\frac{\sqrt {3}}{2}$} \, |\lambda|.
\end{equation}
This shows that the dependence on $\alpha$ is not necessarily continuous.
\end{remark}
We define
    $M \, : \, L^{\infty}(\TT) \rightarrow \cB\big(L^2(\TT)\big),$
    $R \, : \, \TT \rightarrow \cB\big(L^2(\TT)\big),$
by
\begin{eqnarray}
\label{M}
    \big(M(g) \, f\big)(x) &=& g(x)f(x), \qquad g \in L^{\infty}(\TT), \, f \in L^2(\TT), \, x \in \TT, \\
\label{R}
    \big(R(\alpha) \, f\big)(x) &=& f(x + \alpha), \qquad \alpha \in \TT, \, f \in L^2(\TT), \, x \in \TT,
\end{eqnarray}
and let $\cB_{\alpha} \subset \cB\big(L^2(\TT)\big)$ denote the {\it rotation $C^*$--algebra} generated by
the operators $M(\xi_1)$ and $R(\alpha)$ and their adjoints. Then $\big(R(\alpha),M(\xi_1)\big)$ is a frame with parameter $\alpha$ for $\cB_\alpha$, see Definition~\ref{rotdef}.
\begin{pro}
\label{pro:rotrep}
The operators
$H(\alpha,\lambda,\theta),$
$U_H(\kappa,\alpha,\lambda,\theta),$
$U_{kH}(\kappa,\alpha,\lambda,\theta),$ and
$U_{ordkr}(\kappa,\alpha,\lambda,\theta)$
belong to $\cB_{\alpha}$ since
\begin{eqnarray}
H(\alpha,\lambda,\theta) \, &=& \,
    2\Re\big[M(\xi_1)\big]  + 2\lambda\Re \big[\xi_{1}(\theta) \, R(\alpha)\big], \\
U_H(\kappa,\alpha,\lambda,\theta) \, &=& \,
\exp\big[-i2\kappa\Re\big(M(\xi_1)\big) -i2\kappa\lambda\Re \big(\xi_{1}(\theta) \, R(\alpha)\big)\big], \\
U_{kH}(\kappa,\alpha,\lambda,\theta) \, &=& \,
\exp\big[-i2\kappa\Re\big(M(\xi_1)\big)\big] \,
\exp\big[-i2\kappa\lambda\Re\big(\xi_{1}(\theta) \, R(\alpha)\big)\big], \\
U_{ordkr}(\kappa,\alpha,\lambda,\theta) \, &=& \,
\exp\big[-i2\kappa\Re\big(M(\xi_1)\big)\big]\,\exp\big[-i2\kappa\lambda \Re\big(\xi_{1}(\theta+\alpha/2) \, M(\xi_1) \, R(\alpha)\big)\big].
\end{eqnarray}
\end{pro}
{\it Proof.} The first equation follows from $2\Re\left[M(\xi_1)\right]\xi_n = \xi_{n+1} + \xi_{n-1}$
and $2\lambda\Re \left[\xi_{1}(\theta) \, R(\alpha)\right] \xi_n = 2 \lambda \cos\left[ 2\pi(n\alpha+\theta)\right] \, \xi_n.$
The second equation follows from
$\left(-i \alpha \frac{d}{dx} + 2\pi \theta \right)\xi_n = 2\pi(n\alpha + \theta)\xi_n.$
We observe that $T(\alpha) \, \xi_n = \exp\big(i\pi \alpha n^2  \big) \, \xi_n$ and hence the third equation follows from
$T(\alpha) \, M(\xi_1) \, T(\alpha)^{-1}\xi_n = \exp(-i\pi \alpha)\, R(\alpha) \, M(\xi_1) \, \xi_n.$ $\qquad \blacksquare$
\\ \\
We say that $A, B \in
\cB(\HH)$ are {\it unitarily equivalent} (and write $A \cong B$) if
there exists an $O \in \cB_u(\HH)$ such that $A = OBO^{-1}.$ Clearly $A
\cong B$ implies that $\sigma(A) = \sigma(B).$ We say that $A$ and
$B$ are {\it approximately unitarily equivalent} (and write $A
\cong_{a} B$) if there exists a sequence $O_k \in \cB_u(\HH)$ such
that
$\lim_{k \rightarrow \infty} ||A - O_kBO_{k}^{-1}|| = 0.$
Although approximate unitary equivalence is
weaker than unitary equivalence \cite{davidson96},
if $A, B \in \cB_n(\HH)$ and $A \cong_a B$ then $\sigma(A) = \sigma(B).$ This follows since Proposition~\ref{dist} implies that $\sigma \, : \, \cB_n(\HH) \rightarrow \cH(\CC)$ is continuous.
\begin{pro}
\label{pro:specprox}
If $\alpha = p/q$ is rational and $p$ and $q$ are coprime, then for all $\theta_1, \theta_2 \in [0,1),$
\begin{eqnarray}
\label{eq:Hcont2}
d\Big[\sigma \big(H(\alpha,\lambda,\theta_1)\big), \, \sigma \big(H(\alpha,\lambda,\theta_2) \big)\Big] \, & \leq & \,
    2|\lambda| \ \big| \sin[\pi/(2q)] \big|, \\
\label{eq:UHcont2}
d\Big[\sigma \big(U_H(\kappa,\alpha,\lambda,\theta_1)\big), \, \sigma \big(U_H(\kappa,\alpha,\lambda,\theta_2) \big)\Big] \, & \leq & \,
    2|\kappa \, \lambda| \ \big| \sin[\pi/(2q)] \big|, \\
\label{eq:UkHcont2}
d\Big[\sigma \big(U_{kH}(\kappa,\alpha,\lambda,\theta_1)\big), \, \sigma \big(U_{kH}(\kappa,\alpha,\lambda,\theta_2) \big)\Big] \, & \leq & \,
    2|\kappa \, \lambda| \ \big| \sin[\pi/(2q)]  \big|, \\
\label{eq:Uordkrcont2}
d\Big[\sigma \big(U_{ordkr}(\kappa,\alpha,\lambda,\theta_1)\big), \, \sigma \big(U_{ordkr}(\kappa,\alpha,\lambda,\theta_2) \big)\Big] \, & \leq & \,
    2|\kappa \, \lambda| \ \big| \sin[\pi/(2q)] \big|.
\end{eqnarray}
If $\alpha$ is irrational then the spectrums of these operators are independent of $\theta$ and therefore the distances above are zero.
\end{pro}
{\it Proof.} The unitary equivalence
$M(\xi_n) \, R(\alpha) \, M(\xi_n)^{-1} = \xi_1(-n\alpha) \, R(\alpha)$
implies that
${H(\alpha,\lambda,\theta)} \, \cong \, {H(\alpha,\lambda,\theta - n\alpha)},$
$U_H(\kappa,\alpha,\lambda,\theta) \, \cong \, {U_H(\kappa,\alpha,\lambda,\theta - n\alpha)},$
${U_{kH}(\kappa,\alpha,\lambda,\theta)} \, \cong \, {U_{kH}(\kappa,\alpha,\lambda,\theta - n\alpha)}$
and that
${U_{ordkr}(\kappa,\alpha,\lambda,\theta)} \, \cong \, {U_{ordkr}(\kappa,\alpha,\lambda,\theta-n\alpha)}.$
Therefore, Inequality~\ref{eq:Hcont} implies Inequality~\ref{eq:Hcont2}, Inequality~\ref{eq:UHcont} implies Inequality~\ref{eq:UHcont2},
Inequality~\ref{eq:UkHcont} implies Inequality~\ref{eq:UkHcont2}, and
Inequality~\ref{eq:Uordkrcont} implies Inequality~\ref{eq:Uordkrcont2}.
The last statement follows since if $\alpha$ is irrational then $\alpha \ZZ \subset \TT$ is dense
and hence the unitary equivalences above become almost unitary equivalences. $\qquad \blacksquare$
\begin{cor}
\label{UHcantor}
If $\alpha$ is irrational, then the spectrum of $U_{H}(\kappa,\alpha,\lambda,\theta)$ is independent of $\theta$ and is a Cantor set.
\end{cor}
{\it Proof.} Proposition~\ref{pro:specprox} implies the first assertion. Then Lemma~\ref{specmap} implies that
$\sigma\big(U_{H}(\kappa,\alpha,\lambda,\theta)\big)$ is a Cantor set since it equals the image of the Cantor set $\sigma\big(H(\kappa,\alpha,\lambda,\theta)\big)$ under the map $t \rightarrow \exp(-i\kappa t).$ $\qquad \blacksquare$
\\ \\
Following Proposition~\ref{univrot}, for $\alpha \in [0,1)$ we let $(\widetilde {U}_{\alpha}, \widetilde {V}_{\alpha})$ be a frame with parameter $\alpha$ for the universal rotation $C^*$--algebra $\cA_{\alpha}$ and we construct operators, called mother operators, in $\cA_{\alpha}$ by
\begin{eqnarray}
\label{op1}
\widetilde {W}_{\alpha} \, &=& \, \exp(-i\pi \alpha)\, \widetilde {U}_{\alpha}\, \widetilde {V}_{\alpha}, \\
\label{op2}
\widetilde {H}(\alpha,\lambda) \, &=& \, 2\Re \big(\widetilde {V}_{\alpha}\big) + 2\lambda\Re \big(\widetilde {U}_{\alpha}\big), \\
\label{op3}
\widetilde {U}_{H}(\kappa,\alpha,\lambda) \, &=& \, \exp \left[-i2 \kappa \Re \big(\widetilde {V}_{\alpha}\big) -i2\kappa  \lambda\Re \big(\widetilde {U}_{\alpha}\big)\right], \\
\label{op4}
\widetilde {U}_{kH}(\kappa,\alpha,\lambda) \, &=& \,
    \exp \left[-i2 \kappa \Re \big(\widetilde {V}_{\alpha}\big)\right] \, \exp \left[-i2 \kappa \lambda\Re \big(\widetilde {U}_{\alpha}\big)\right], \\
\label{op5}
\widetilde {U}_{ordkr}(\kappa,\alpha,\lambda) \, &=& \,
    \exp \left[-i2 \kappa \Re \big(\widetilde {V}_{\alpha}\big)\right] \, \exp \left[-i2 \kappa \lambda \Re \big(\widetilde {W}_{\alpha}\big)\right].
\end{eqnarray}
\begin{theorem}
\label{main1}
The families of operators defined by the equations above satisfy
\begin{eqnarray}
\label{theta1}
\pi_{\theta}\big(\widetilde {H}(\alpha,\lambda)\big) \, &=& \, H(\alpha,\lambda,\theta), \\
\label{theta2}
\pi_{\theta}\big(\widetilde {U}_{H}(\kappa,\alpha,\lambda)\big) \, &=& \, U_{H}(\kappa,\alpha,\lambda,\theta), \\
\label{theta3}
\pi_{\theta}\big(\widetilde {U}_{kH}(\kappa,\alpha,\lambda)\big) \, &=& \, U_{kH}(\kappa,\alpha,\lambda,\theta), \\
\label{theta4}
\pi_{\theta}\big(\widetilde {U}_{ordkr}(\kappa,\alpha,\lambda)\big) \, &=& \, U_{ordkr}(\kappa,\alpha,\lambda,\theta),
\end{eqnarray}
where $\pi_\theta, \, \theta \in \TT \in [0,1),$ is the family of homomorphisms constructed in Lemma~\ref{pitheta},
and their spectrums satisfy
\begin{eqnarray}
\label{sigmafirst}
\sigma\big(\widetilde {H}(\alpha,\lambda)\big) \, &=& \, \bigcup_{\theta \in [0,\frac{1}{q})}\sigma\big(H(\alpha,\lambda,\theta)\big), \\
\label{sigmasecond}
\sigma\big(\widetilde {U}_{H}(\kappa,\alpha,\lambda)\big) \, &=&
\, \bigcup_{\theta \in [0,\frac{1}{q})} \sigma\big(U_{H}(\kappa,\alpha,\lambda,\theta)\big), \\
\label{sigmamiddle}
\sigma\big(\widetilde {U}_{kH}(\kappa,\alpha,\lambda)\big) \, &=&
\, \bigcup_{\theta \in [0,\frac{1}{q})} \sigma\big(U_{kH}(\kappa,\alpha,\lambda,\theta)\big), \\
\label{sigmalast}
\sigma\big(\widetilde {U}_{ordkr}(\kappa,\alpha,\lambda)\big) \, &=&
\, \bigcup_{\theta \in [0,\frac{1}{q})} \sigma\big(U_{ordkr}(\kappa,\alpha,\lambda,\theta)\big)
\end{eqnarray}
and are continuous functions of $\kappa, \alpha,$ and $\lambda$
since
\begin{eqnarray}
\label{ineq1}
d\left[ \sigma\big(\widetilde {H}(\alpha_1,\lambda)\big), \,
\sigma\big(\widetilde {H}(\alpha_2,\lambda)\big) \right] \, &\leq& \,
36 \, \sqrt {6\pi |\lambda(\alpha_2-\alpha_1)|}\, , \\
\label{ineq2}
d\left[ \sigma\big(\widetilde {U}_{H}(\kappa,\alpha_1,\lambda)\big), \,
\sigma\big(\widetilde {U}_{H}(\kappa,\alpha_2,\lambda)\big) \right] \, &\leq&
\, 36 \, \sqrt {6\pi |\kappa \lambda(\alpha_2-\alpha_1)|}\, , \\
\label{ineq3}
d\left[ \sigma\big(\widetilde {U}_{kH}(\kappa,\alpha_1,\lambda)\big), \,
\sigma\big(\widetilde {U}_{kH}(\kappa,\alpha_2,\lambda)\big) \right] \, &\leq&
\, 36 \, \sqrt {6\pi |\kappa \lambda(\alpha_2-\alpha_1)|}\, , \\
\label{ineq4}
d\left[ \sigma\big(\widetilde {U}_{ordkr}(\kappa,\alpha_1,\lambda)\big), \,
\sigma\big(\widetilde {U}_{ordkr}(\kappa,\alpha_2,\lambda)\big) \right] \, &\leq&
\, 36 \, \sqrt {6\pi |\kappa \lambda(\alpha_2-\alpha_1)|}\,
\end{eqnarray}
Furthermore, there exists an $L \in \cB\big(L^2(\TT^2)\big)$ such that
\begin{equation}
\label{con}
    \widetilde {U}_{kH}(\kappa,\alpha,\lambda) \ \cong
        \ L \, \widetilde {U}_{kH}(\kappa,\alpha,\lambda) \, L^{-1} \ = \ \widetilde {U}_{ordkr}(\kappa,\alpha,\lambda)
\end{equation}
and hence
$\sigma \big( \widetilde {U}_{kH}(\kappa,\alpha,\lambda)\big) \, =
\, \sigma \big( \widetilde {U}_{ordkr}(\kappa,\alpha,\lambda) \big).$
\end{theorem}
{\it Proof.} Proposition~\ref{pro:rotrep} and Lemma~\ref{pitheta} implies that the homomorphism $\pi_\theta$ maps the operators
defined by Equations \ref{op2}--\ref{op5} onto the operators on the right side of Equations \ref{theta1}--\ref{theta4}.
The proof of Proposition~\ref{pro:specprox} shows that each of the operators that appear on the right in Equations \ref{sigmafirst}--\ref{sigmalast} is unitarily equivalent to itself with $\theta$ replaced by $\theta + 1/q.$ Therefore, Proposition~\ref{separate} implies Equations \ref{sigmafirst}--\ref{sigmalast}.
Let $\mu > 0$ and let $\pi_j \, : \, \cA_{\alpha_j} \rightarrow \cB(\HH)$ for $j = 1,2$ be the injective homomorphisms described in Proposition~\ref{embed}. Define operators $H_j \in \cB(\HH),
\, j = 1,2$ by
$
    H_j = \pi_j \big( \widetilde {H}(\alpha_j,\lambda)\big).
$
Since $\pi_j$ is injective
$\sigma\big(\widetilde {H}(\alpha_j,\lambda)\big) = \sigma(H_j)$
and hence Inequality~\ref{eq:speccont} implies that
$d\big[ \sigma\big(\widetilde {H}(\alpha_1,\lambda)\big), \,
\sigma\big(\widetilde {H}(\alpha_2,\lambda)\big) \big] = d\big[\sigma(H_2), \, \sigma(H_1) \big] \leq ||H_2-H_1||.$
The triangle inequality together with Inequalities \ref{ineqalpha12} implies that
$$||H_2-H_1|| \ \leq \ 2 \big|\big|\Re \pi_2(\widetilde {V}_{\alpha_2}) - \Re \pi_1(\widetilde {V}_{\alpha_1})\big|\big| +
2|\lambda| \, \big|\big|\Re \pi_2(\widetilde {U}_{\alpha_2}) - \Re \pi_1(\widetilde {U}_{\alpha_1}) \big|\big|
\ \leq \ \frac{54}{\mu} + 18 |\lambda| \, \mu.$$
Choosing $\mu = \sqrt {3/|\lambda|}$ minimizes the right side of the inequality above and gives Inequality~\ref{ineq1}. Inequalities \ref{ineq2}--\ref{ineq4} are obtained using the same procedure.
Proposition~\ref{auto} implies Equation~\ref{con} and completes the proof. $\qquad \blacksquare$
\section{Algorithms and Numerical Examples}
\label{algorithms}
\begin{figure}[p]  
\includegraphics[width=0.5\textwidth, viewport = 140 80 410 350,clip]{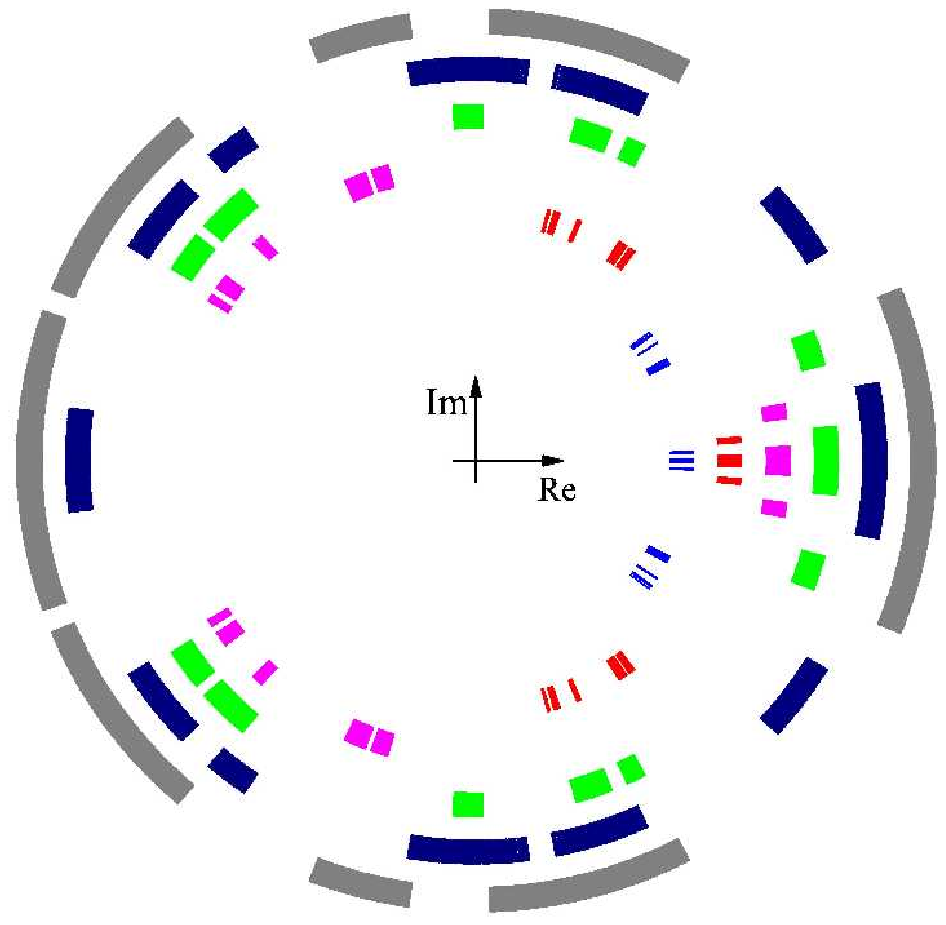}
\caption{Spectrums for mother unitary Harper operators, $\widetilde {U}_{H}(\kappa,\alpha,\lambda),$ for $\alpha = 8/13$ and $\lambda = 1.$ As stated in detail at the end of Section~\ref{algorithms}, each spectrum is a subset of the unit circle. The figure plots several of these spectrums on concentric circles whose radii increase with $\kappa \in \{0.25, 0.5, 1, 2, 4, 8\}.$ The real and imaginary axes are indicated at the common center of the circles. These estimated spectrums are computed using a grid of $100$ points in both $x$ and $\theta$, using Equation~\ref{UHxtheta}. \label{fig:Harp}}
\end{figure}

\begin{figure}[p]
\includegraphics[width=0.5\textwidth, viewport = 140 80 410 350,clip]{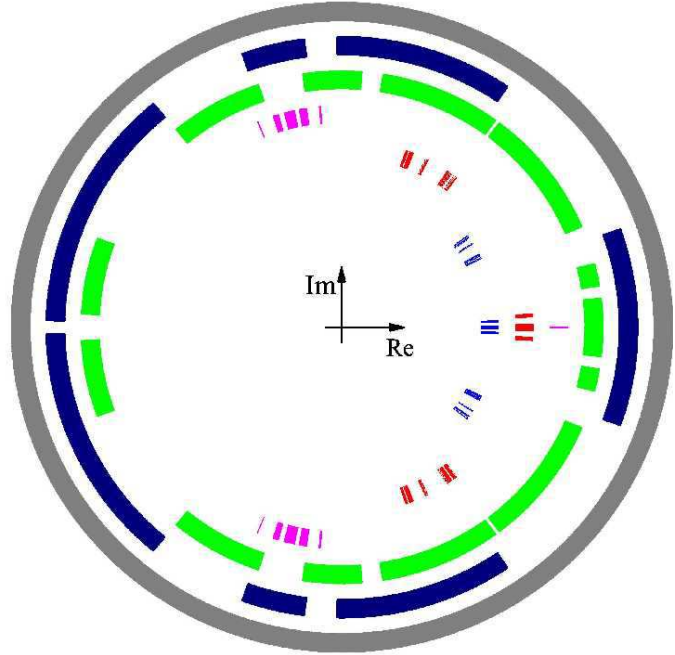}
\caption{Spectrums for mother kicked Harper operators, $\widetilde {U}_{kH}(\kappa,\alpha,\lambda),$ using the same parameters as in Figure~\ref{fig:Harp}, and using Equation~\ref{UkHxtheta} for computations, also with the same $(x,\theta)$--grid as in Figure~\ref{fig:Harp}. Comparing the spectrums in this figure with those in Figure~\ref{fig:kRot} clearly illustrates Theorem~\ref{main1}, namely that the spectrums are identical.  \label{fig:kHarper}}
\end{figure}

\begin{figure}[p]
\includegraphics[width=0.5\textwidth, viewport = 140 80 410 350,clip]{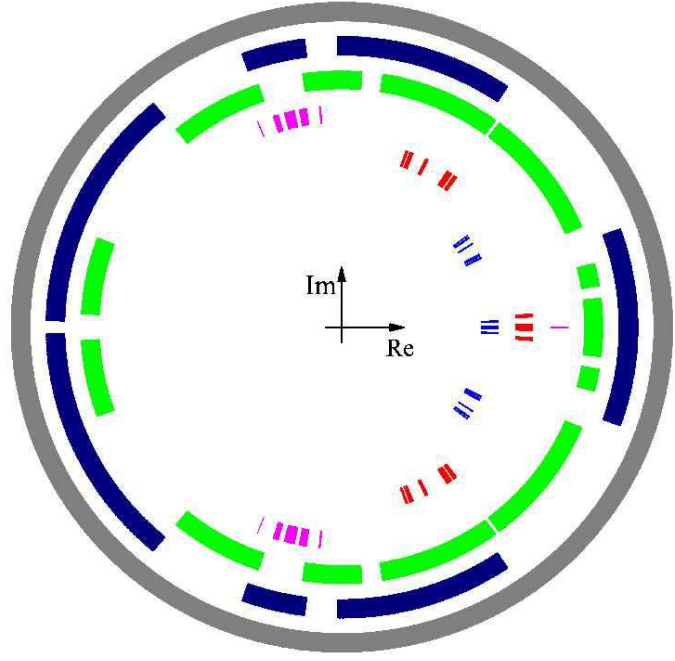}
\caption{Spectrums for the mother on--resonance double kicked rotor operators, $\widetilde {U}_{ordkr}(\kappa,\alpha,\lambda),$ using the same parameters as in Figure~\ref{fig:Harp} and the same $(x,\theta)$--grid for computations. Theorem~\ref{main1} says that the spectrums in this figure are identical with those in Figure~\ref{fig:kHarper}. \label{fig:kRot}}
\end{figure}

\begin{figure}[p]
\includegraphics[width=0.5\textwidth, viewport = 140 80 410 350,clip]{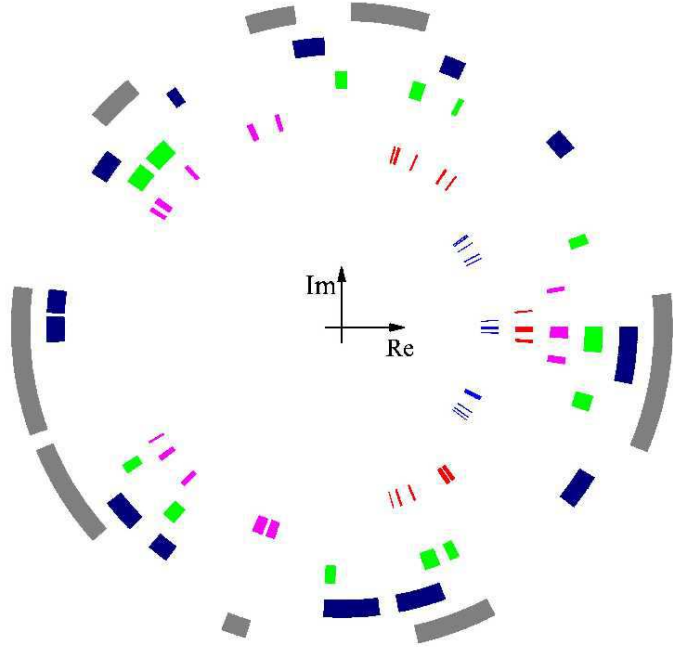}
\caption{Spectrums for unitary Harper operators, ${U}_{H}(\kappa,\alpha,\lambda,\theta),$ for $\alpha = 8/13, \, \lambda = 1, \theta = 0$ and for several values of $\kappa \in \{0.25, 0.5, 1, 2, 4, 8\}.$ As stated in detail at the end of Section~\ref{algorithms}, each spectrum is a subset of the unit circle. The figure plots several of these spectrums on concentric circles whose radii increase with $\kappa.$ The real and imaginary axes are indicated at the common center of the circles. We note that these spectrums are proper subsets of the physically more relevant spectrums displayed in Figure~\ref{fig:Harp}. These estimated spectrums are computed using a grid of $10,000$ points in $x$, using Equation~\ref{UHx}. \label{fig:Harper1theta}}
\end{figure}

\begin{figure}[p]
\includegraphics[width=0.5\textwidth, viewport = 140 80 410 350,clip]{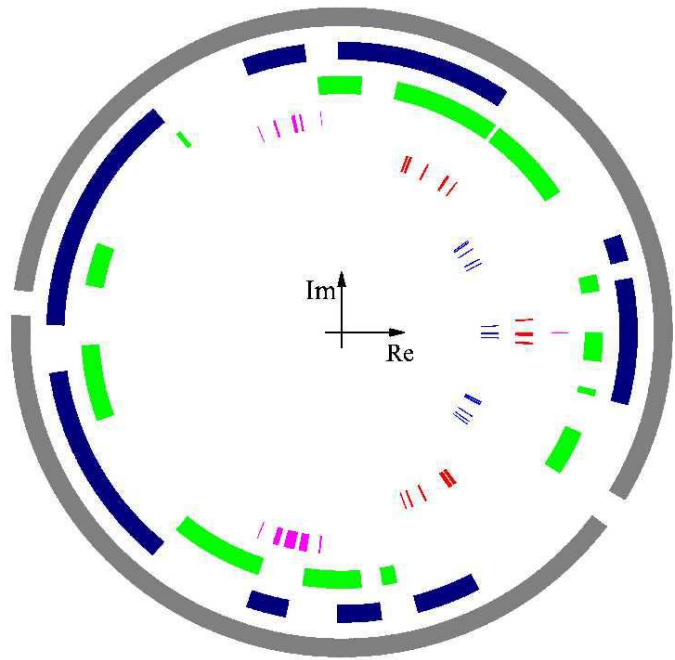}
\caption{Spectrums for kicked Harper operators, ${U}_{kH}(\kappa,\alpha,\lambda,\theta),$ using the same parameters as in Figure~\ref{fig:Harper1theta}, and using Equation~\ref{UkHx} for computations, also with the same $x$--grid as in Figure~\ref{fig:Harper1theta}. The spectrums shown here are proper subsets of those shown in Figure~\ref{fig:kHarper}. Compare these spectrums with the spectrums for the on--resonance double kicked Harper operators, also for $\theta = 0,$ shown in Figure~\ref{fig:kRot1theta}. \label{fig:kHarper1theta}}
\end{figure}

\begin{figure}[p]
\includegraphics[width=0.5\textwidth, viewport = 140 80 410 350,clip]{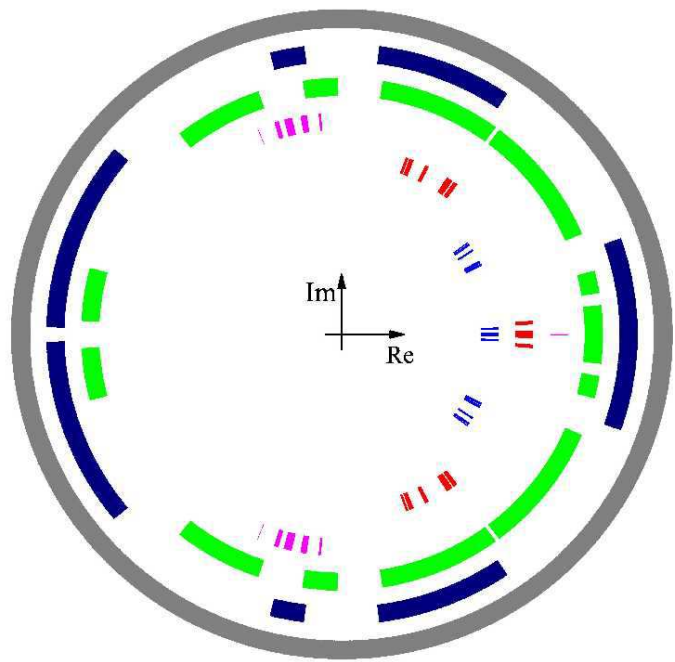}
\caption{Spectrums for the on--resonance double kicked rotor operators, ${U}_{ordkr}(\kappa,\alpha,\lambda,\theta),$ using the same parameters as in Figure~\ref{fig:Harper1theta}. The spectrums shown here are proper subsets of those shown in Figure~\ref{fig:kRot}. In contrast to the spectrums of the mother on--resonance double kicked rotor operators shown in Figure~\ref{fig:kRot}, which are identical to those of the mother kicked Harper operators shown in Figure~\ref{fig:kHarper}, these spectrums differ from the spectrums of the kicked Harper operators for fixed $\theta = 0$ shown in Figure~\ref{fig:kHarper1theta}. \label{fig:kRot1theta}}
\end{figure}

Assume that $\alpha = p/q,$ where $p \in \ZZ,$ $q \in \NN,$ with $p$ and $q$ coprime.
We will derive algorithms that accurately estimate the spectrums of the unitary operators
$U_{H}(\kappa,\alpha,\lambda,\theta),$
$U_{kH}(\kappa,\alpha,\lambda,\theta)$
and
$U_{ordkr}(\kappa,\alpha,\lambda,\theta)$
and of their mother unitary operators
$\widetilde {U}_{H}(\kappa,\alpha,\lambda),$
$\widetilde {U}_{kH}(\kappa,\alpha,\lambda)$
and
$\widetilde {U}_{ordkr}(\kappa,\alpha,\lambda).$
The algorithms work by computing eigenvalues of matrices in $\cM_q,$ the set of $q \times q$ matrices. The matrices are parameterized by $(x,\theta) \in [0,1/q)^2$
and the estimates are obtained by forming the union of the sets of eigenvalues over an equal--spaced $N \times N$ grid of values of $(x,\theta).$
\\ \\
Let $\omega = \exp(i 2 \pi /q)$ and let $F, C,$ and $D$ be the $q \times q$ matrices defined by Equations \ref{FCD}.
For $x \in [0,1)$ construct the homomorphism $\rho(x) \in \textnormal{Hom}(\cB_\alpha,\cM_q)$ so that
$\rho(x)\big(M(\xi_1)\big) = \xi_1(x)D$ and $\rho(x)(R_{\alpha}) = C^p.$ Proposition~\ref{irredrep3} ensures the existence of this
homomorphism. For any integer $k$ and $y \in [0,1)$ define the function
$c(y) = \cos(2 \pi y)$ and define the $q \times q$ matrix
$$
    G(k,y) = \left[
    \begin{array}{cccc}
      c(y) &  &  &  \\
       & c\big(y+\frac{k}{q}\big) &  &  \\
       &  & \ddots &  \\
       &  &  & c\big(y+\frac{k(q-1)}{q}\big)
    \end{array} \right].
$$
The following result provides the basis for the algorithms for the unitary Harper and unitary kicked Harper operators.
\begin{theorem}
\label{algH}
For $\alpha = p/q,$ $\, x, \theta \in [0,1)$ and $\kappa, \lambda \in \RR$ we define the matrices
\begin{equation}
\label{MUH}
  M_{U_H}(\kappa,\alpha,\lambda,x,\theta) \, = \,
 \exp\Big[-i2\kappa \, \big( G(1,x) \, + \lambda \, F \, G(p,\theta) \, F^{-1} \big) \Big],
\end{equation}
and
\begin{equation}
\label{MUkH}
  M_{U_{kH}}(\kappa,\alpha,\lambda,x,\theta) \, = \,
 \exp\big[-i2\kappa \, G(1,x)\big] \, F \, \exp\big[-i2\kappa \lambda \, G(p,\theta)\big] \, F^{-1}
\end{equation}
$$
\cong
  F^{-1}\left[
    \begin{array}{cccc}
      e^{-i2\kappa \, c(x)} &  &  &  \\
       &  e^{-i2\kappa \, c\big(x+\frac{1}{q}\big)} &  &  \\
       &  & \ddots &  \\
       &  &  & e^{-i2\kappa \, c\big(x+\frac{q-1}{q}\big)}
    \end{array} \right] F \, \left[ \begin{array}{cccc}
      e^{-i2\kappa \lambda \, c(\theta)} &  &  &  \\
       &  e^{-i2\kappa \lambda \, c\big(\theta+\frac{p}{q}\big)} &  &  \\
       &  & \ddots &  \\
       &  &  & e^{-i2\kappa \lambda \, c\big(\theta+\frac{p(q-1)}{q}\big)}
    \end{array} \right]
$$
$$
=
  \left[
    \begin{array}{ccccc}
      v_1 &  v_2 & \cdots & v_{q-1} & v_q \\
      v_q &  v_1 & \cdots & v_{q-2} & v_{q-1} \\
   \vdots & \vdots & \ddots &  \vdots & \vdots \\
      v_3 & v_4 & \cdots  & v_1 & v_2  \\
      v_2 & v_3 & \cdots &  v_q & v_1
    \end{array} \right] \, \left[ \begin{array}{ccccc}
      e^{-i2\kappa \lambda \, c(\theta)} &  &  &  & \\
       &  e^{-i2\kappa \lambda \, c\big(\theta+\frac{p}{q}\big)} &  &  & \\
       &  & \ddots & & \\
       &  & & e^{-i2\kappa \lambda \, c\big(\theta+\frac{(q-2)p}{q}\big)} & \\
       &  & & & e^{-i2\kappa \lambda \, c\big(\theta+\frac{(q-1)p}{q}\big)}
    \end{array} \right],
$$
where
$$
\left[v_1 \ \ v_2 \ \ v_3 \ \ \cdots \ \ v_q\right]^T \ = \ \frac{1}{\sqrt {q}} \, F \,
\left[e^{-i2\kappa \, c(x)} \ \ e^{-i2\kappa \, c\big(x+\frac{1}{q}\big)} \ \  e^{-i2\kappa \, c\big(x+\frac{2}{q}\big)} \cdots \ \ e^{-i2\kappa \lambda \, c\big(x+\frac{(q-1)}{q}\big)}\right]^T.
$$
If $\alpha = p/q,$ $\, \theta \in [0,1)$ and $\kappa, \lambda \in \RR,$ then
\begin{equation}
\label{UHx}
\sigma\big(U_{H}(\kappa,\alpha,\lambda,\theta)\big) \, = \, \bigcup_{x \in [0,\frac{1}{q})} \sigma\big(M_{U_H}(\kappa,\alpha,\lambda,x,\theta)\big),
\end{equation}
and
\begin{equation}
\label{UkHx}
\sigma\big(U_{kH}(\kappa,\alpha,\lambda,\theta)\big) \, = \, \bigcup_{x \in [0,\frac{1}{q})} \sigma\big(M_{U_{kH}}(\kappa,\alpha,\lambda,x,\theta)\big).
\end{equation}
If a uniformly spaced grid of $N \in \NN$ values of $x$ is used to compute an estimate
$\sigma_N\big({U}_{kH}(\kappa,\alpha,\lambda,\theta)\big)$ then
\begin{equation}
\label{estUHx}
d\Big[\sigma_N\big({U}_{H}(\kappa,\alpha,\lambda,\theta)\big), \, \sigma\big({U}_{H}(\kappa,\alpha,\lambda,\theta)\big)\Big] \
\leq \ \frac{2\pi |\kappa|}{Nq},
\end{equation}
and
\begin{equation}
\label{estUkHx}
d\Big[\sigma_N\big({U}_{kH}(\kappa,\alpha,\lambda,\theta)\big), \, \sigma\big({U}_{kH}(\kappa,\alpha,\lambda,\theta)\big)\Big] \
\leq \ \frac{2\pi |\kappa|}{Nq}.
\end{equation}
If $\alpha = p/q$ and $\kappa, \lambda \in \RR,$ then
\begin{equation}
\label{UHxtheta}
\sigma\big(\widetilde {U}_{H}(\kappa,\alpha,\lambda)\big) \, = \, \bigcup_{x \in [0,\frac{1}{q})} \ \bigcup_{\theta \in [0,\frac{1}{q})} \sigma\big(M_{U_H}(\kappa,\alpha,\lambda,x,\theta)\big),
\end{equation}
and
\begin{equation}
\label{UkHxtheta}
\sigma\big(\widetilde {U}_{kH}(\kappa,\alpha,\lambda)\big) \, = \, \bigcup_{x \in [0,\frac{1}{q})} \ \bigcup_{\theta \in [0,\frac{1}{q})}\sigma\big(M_{U_{kH}}(\kappa,\alpha,\lambda,x,\theta)\big).
\end{equation}
Furthermore, if a uniformly spaced two-dimensional grid of $N \times N$ values of $(x,\theta) \in [0,\frac{1}{q})^2$ is used to compute an estimate
$\sigma_{N \times N}\big(\widetilde {U}_{kH}(\kappa,\alpha,\lambda)\big)$ then
\begin{equation}
\label{estUHxtheta}
d\Big[\sigma_{N \times N}\big(\widetilde {U}_{H}(\kappa,\alpha,\lambda)\big),\, \sigma\big(\widetilde {U}_{H}(\kappa,\alpha,\lambda)\big)\Big] \ \leq \ \frac{2\pi |\kappa| (1+|\lambda|)}{Nq},
\end{equation}
and
\begin{equation}
\label{estUkHxtheta}
d\Big[\sigma_{N \times N}\big(\widetilde {U}_{kH}(\kappa,\alpha,\lambda)\big),\, \sigma\big(\widetilde {U}_{kH}(\kappa,\alpha,\lambda)\big)\Big] \ \leq \ \frac{2\pi |\kappa| (1+|\lambda|)}{Nq}.
\end{equation}
\end{theorem}
{\it Proof.} Since $\rho(x)(R_{\alpha}) \, = \, C^p \, = \, F \, D^{p} \, F^{-1}$ it follows that
$\rho(x)\big(U_{H}(\kappa,\alpha,\lambda,\theta)\big) \, = \, M_{U_{H}}(\kappa,\alpha,\lambda,\theta)$
and
$\rho(x)\big(U_{kH}(\kappa,\alpha,\lambda,\theta)\big) \, = \, M_{U_{kH}}(\kappa,\alpha,\lambda,\theta).$
Therefore, since $C \, G(1,x) \, C^{-1} \, = \, G(1,x+1/q),$ Proposition~\ref{separate} implies
Equations \ref{UHx} and \ref{UkHx}. Furthermore, Equations \ref{sigmasecond} and \ref{sigmamiddle} imply
Equations \ref{UHxtheta} and \ref{UkHxtheta}.
Inequalities~\ref{estUHx} and \ref{estUkHx} are derived from
Equation~\ref{eq:speccont}, using Equation~\ref{eq:expcont} to show that
$$\big|\big|M(\kappa,\alpha,\lambda,x_1,\theta) - M(\kappa,\alpha,\lambda,x_2,\theta)\big|\big| \ \leq \ 2|\kappa| \, \big|\big|G(1,x_1)-G(1,x_2)\big|\big| \ \leq \ 4\pi|\kappa(x_2-x_1)|,$$
and the fact that the distance between every $x \in [0,1/q)$ and a uniform grid with $N$ points in $[0,1/q)$ is $\leq 1/(2Nq).$
Inequalities~\ref{estUHxtheta} and \ref{estUkHxtheta} are derived likewise from Equation~\ref{eq:speccont} using Equation~\ref{eq:expcont} to show that
\begin{eqnarray}
\big|\big|M(\kappa,\alpha,\lambda,x_1,\theta_1) - M(\kappa,\alpha,\lambda,x_2,\theta_2)\big|\big| \
&\leq& \ 2|\kappa| \, \big|\big|G(1,x_1)-G(1,x_2)\big|\big| + 2|\kappa \lambda| \, \big|\big|G(p,\theta_1)-G(p,\theta_2)\big|\big| \nonumber \\
\ &\leq& \ 2\pi|\kappa(x_1-x_2)| + 2\pi|\kappa\lambda(\theta_1-\theta_2)|. \nonumber
\end{eqnarray}
This concludes the proof. $\qquad \blacksquare$
\\ \\
We now consider on--resonance double kicked rotor operators. Clearly,
\begin{equation}
\label{rhoxor1}
\rho(x)\big(U_{ordkr}(\kappa,\alpha,\lambda,\theta)\big) \, = \,
\exp\Big[-i2\kappa \, G(1,x)\Big] \, \exp\Big[-i 2\kappa \lambda \, \Re\big( \, \xi_{1}(\theta+\alpha/2) \, \xi_1(x) \, D \, C^p \big) \, \Big].
\end{equation}
In order to develop an efficient algorithm we need to diagonalize the matrix $DC^p.$ Let $\nu \in \CC$ be an eigenvalue
of $DC^p.$ Then there exists a (nonzero) eigenvector $u = [u_1 \ u_2 \ \cdots \ u_q]^T$ such that $DC^p u = \nu u.$ We consider
the indices of $u$ modulo $q,$ compute
\begin{eqnarray}
  u_{p+1} \, &=&  \, \nu \, u_1, \nonumber\\
  u_{2p+1} \, &=& \, \omega^{-p} \, \nu^2 \, u_1, \nonumber\\
  u_{3p+1} \, &=& \, \omega^{-p-2p} \, \nu^3 \, u_1, \nonumber\\
  u_{4p+1} \, &=& \, \omega^{-p-2p-3p} \, \nu^4 \, u_1, \nonumber\\
  &\vdots&  \nonumber\\
\label{nunu}
  u_{qp+1} \, &=& \, \omega^{-p-2p-3p-\ldots-(q-1)p} \, \nu^q \, u_1 \ = \ u_{1} \, ,
\end{eqnarray}
and then observe that $DC^pu = \nu u$ if and only if $\omega^{-pq(q-1)/2} \, \nu^q = 1$ and the entries of $u$ satisfy the equations above. Clearly, $\omega^{-pq(q-1)/2} = (-1)^{p(q-1)}$ so the eigenvalues of $DC^p$ equal $\nu_k = \mu \omega^{k-1}$ with $k = 1,2,...,q,$ and where $\mu = 1$ if $p(q-1)$ is even and $\mu = e^{i\pi/q}$ if $p(q-1)$ is odd. Let $\Lambda = \textnormal{diag}(\nu_1,\nu_2,...,\nu_q)$ and let
$E$ be the $q \times q$ matrix whose columns are the corresponding normalized eigenvectors, computed using Equations \ref{nunu}, so that $DC^pE = E\Lambda.$ Then Equation~\ref{rhoxor1} implies that
\begin{equation}
\label{rhoxor2}
\rho(x)\big(U_{ordkr}(\kappa,\alpha,\lambda,\theta)\big) =
\exp\big[-i2\kappa \, G(1,x) \big] \, E \, \exp\big[-i2\kappa \lambda \, \Re\big(\, \xi_{1}(\theta+\alpha/2) \, \xi_1(x)\, \Lambda \, \big)  \, \big] \, E^{-1}
\end{equation}
$$
= \left[
    \begin{array}{cccc}
      e^{-i2\kappa \, c(x)} &  &  &  \\
       &  e^{-i2\kappa \, c\big(x+\frac{1}{q}\big)} &  &  \\
       &  & \ddots &  \\
       &  &  & e^{-i2\kappa \, c\big(x+\frac{q-1}{q}\big)}
    \end{array} \right] E \left[ \begin{array}{cccc}
      e^{-i2\kappa \lambda \, c(\beta)} &  &  &  \\
       &  e^{-i2\kappa \lambda \, c\big(\beta + \frac{1}{q}\big)} &  &  \\
       &  & \ddots &  \\
       &  &  & e^{-i2\kappa \lambda \, c\big(\beta + \frac{(q-1)}{q}\big)}
    \end{array} \right]E^{-1}
$$
where $\beta = x + \theta + \alpha/2 + \phi$ and $\mu = e^{i2\pi \phi}.$ Equations and Inequalities completely analogous to those in Theorem~\ref{algH} can be derived for the on--resonance double kicked rotor operators.
\\ \\
To illustrate these results, Figures~\ref{fig:Harp}--\ref{fig:kRot1theta} show examples of the various operator's spectrums. Figure~\ref{fig:Harp} illustrates $\sigma\big(\widetilde {U}_{H}(\kappa,\alpha,\lambda)\big),$
Figure~\ref{fig:kHarper} illustrates $\sigma\big(\widetilde {U}_{kH}(\kappa,\alpha,\lambda)\big),$
and
Figure~\ref{fig:kRot} illustrates $\sigma\big(\widetilde {U}_{ordkr}(\kappa,\alpha,\lambda)\big),$
for the following parameter values: $\kappa = 0.25, \ 0.5, \ 1, \ 2, \ 4, \ 8,$ $ \ \alpha = 8/13,$ and $ \, \lambda = 1.$
A grid of $100$ values of both $x$ and $\theta$ were used for each $\kappa$ to compute estimates for each of these spectrums. Thus, each estimate consists of a subset (having $100 \times 100 \times 13 = 130,000$ points) of the spectrum. Since the operators are unitary, their spectrums are subsets of the unit circle in the complex plane, and we have plotted the estimated spectrums on circles to reflect this geometry. However, we plot the estimated spectrums on circles whose radii increase with $\kappa$ to enable them to be compared visually.
We recall from Section~\ref{intro} that the spectrum of the almost Mathieu mother operator $\sigma\big(\widetilde {H}(\kappa,\alpha,\lambda)\big)$ consists of $q$ disjoint intervals if $q$ is odd and consists of $q-1$ disjoint intervals if $q$ is even.
Since $\widetilde {U}_{H}(\kappa,\alpha,\lambda) = \exp\big(-i\kappa \widetilde {H}(\alpha,\lambda) \big),$
the Spectral Mapping Lemma~\ref{specmap} implies that $\sigma\big(\widetilde {U}_{H}(\kappa,\alpha,\lambda)\big)$ equals the image of the union of these intervals under the map $t \rightarrow \exp(-i\kappa t).$ Therefore, for sufficiently small values of $|\kappa|,$ the spectrum
$\sigma\big(\widetilde {U}_{H}(\kappa,\alpha,\lambda)\big)$ also consists of $q$ disjoint intervals if $q$ is odd and consists of $q-1$ disjoint intervals if $q$ is even and the lengths of each of these intervals is proportional to $|\kappa|.$ As $|\kappa|$ increases, some of these intervals will begin to overlap and eventually each will wrap around the unit circle and the spectrum will equal the entire unit circle. These tendencies are clearly illustrated in Figure~\ref{fig:Harp}.
A comparison of Figures~\ref{fig:kHarper} and \ref{fig:kRot} shows that the spectrums of the mother kicked operators (for identical parameter values) are identical, thereby illustrating Theorem~\ref{main1}. Moreover, it can be shown, using the Baker-Campbell-Hausdorff formula and operator splitting, that the Hausdorff distance between these spectrums and $\sigma\big(\widetilde {U}_{H}(\kappa,\alpha,\lambda)\big)$ decreases with $O(\kappa^3),$ which can be assessed from comparing the figures.
Figure~\ref{fig:Harper1theta} illustrates $\sigma\big({U}_{H}(\kappa,\alpha,\lambda,\theta)\big),$
Figure~\ref{fig:kHarper1theta} illustrates $\sigma\big({U}_{kH}(\kappa,\alpha,\lambda,\theta)\big),$
and
Figure~\ref{fig:kRot1theta} illustrates $\sigma\big({U}_{ordkr}(\kappa,\alpha,\lambda,\theta)\big),$
for the following parameter values: $\kappa = 0.25, \ 0.5, \ 1, \ 2, \ 4, \ 8,$ $ \ \alpha = 8/13, \lambda = 1$ and fixed $\theta = 0.$
In these Figures~\ref{fig:Harper1theta}--\ref{fig:kRot1theta}, $10,000$ values of $x$ were used to calculate each estimate for the spectrums, yielding the same total number of points in each spectrum estimate as in the other figures.
For a fixed value of $\theta$, the spectrums are proper subsets of the spectrums of their respective mother operators. Even though the spectrums for their associated mother operators are identical, the spectrums of $\sigma\big({U}_{kH}(\kappa,\alpha,\lambda,\theta)\big)$ and $\sigma\big({U}_{ordkr}(\kappa,\alpha,\lambda,\theta)\big)$ for a fixed value of $\theta$ are different, as is apparent by comparing Figures~\ref{fig:kHarper1theta} and \ref{fig:kRot1theta}.
\\ \\
\section{Future Research}
\label{sec:futureresearch}
\begin{figure}[p]
\includegraphics[width = 0.5\textwidth, viewport = 70 60 470 350,clip]{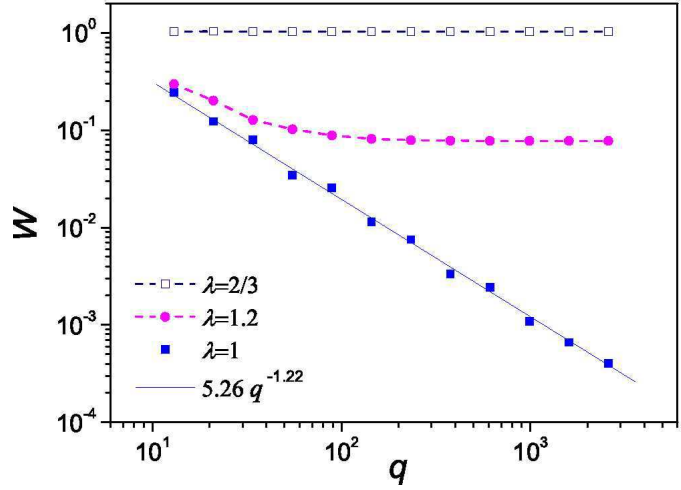}
\caption{The figure shows a log--log plot of the combined width, $W$, of all spectral bands as a function of $q$ for three different values of $\lambda = 2/3, 1$ and $1.2$. In these calculations, $\kappa = 1$ and the error for each data point is smaller than $10^{-9}$. Interestingly, the combined width for $\lambda = 1$ seems to approximately follow a power--law in $q$. A fit to these data--points yields the fitting parameters $5.26 \pm 0.66$ for the pre--exponential factor and $-1.22 \pm 0.02$ for the exponent. This result suggests that the spectrum's measure for $\lambda = 1$ and irrational $\alpha$ could be zero. \label{fig:allwidths}}
\end{figure}

\begin{figure}
\includegraphics[width = 0.5\textwidth, viewport = 90 90 470 350,clip]{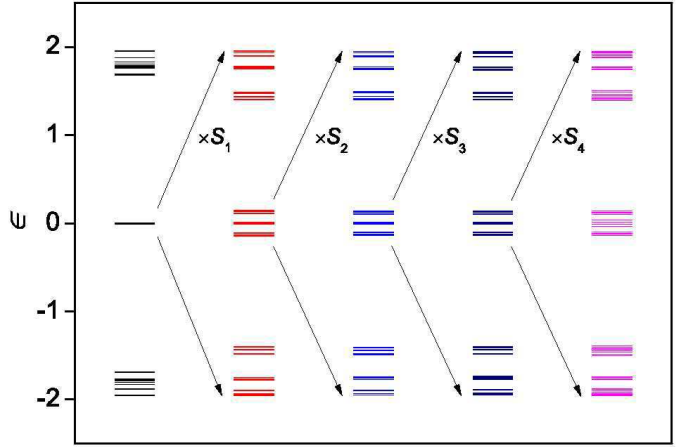}
\caption{The figures shows $\epsilon$, the imaginary part of the logarithm of the spectrum of $\widetilde U_{kH}(\kappa,\alpha,\lambda)$ for $\kappa = \lambda = 1$ and $\alpha = p/q = 2584/4181$. This value of $\alpha$ is a ratio of Fibonacci numbers approximating the reciprocal $(\sqrt {5} - 1)/2$ of the Golden Mean $(\sqrt {5} + 1)/2$, obtained by truncating its continued fraction expansion. The first column shows the whole spectrum, and the vertical axis indicates its values of $\epsilon$. The second column is a zoom-in of the first column's center part, with the zoom-in factor $S_1 = 290$. The third to the fifth columns are the zoom-ins of the column preceding them with zoom--in factors of $S_2 = 13.3$ and $S_3 = S_4 = 14$. There seems to be considerable self--similarity, especially between the second to fourth column. The bands of the fifth column are wider than their predecessors, due to the finiteness of $q$, which indicates that we have reached the limit of finding self--similarity by zooming in. \label{fig:UkHzoom}}
\end{figure}

One class of future research projects is to improve existing algorithms and to develop new algorithms for computing spectrums. For rational $\alpha,$ a direct method that estimates the spectral bands by directly computing their endpoints would be more efficient than our current method that forms unions of sets of eigenvalues of matrices over a grid of values of $x$ and $\theta.$ Moreover, new algorithms for computing global topological properties of the eigenvalues, such as the Chern numbers computed by Leboeuf {\it et al.} in \cite{leboeuf90}, are desirable.
\\ \\
Another class of projects will address the physical significance of spectral characteristics. These include conductivity,
diffusion and other transport phenomena.
\\ \\
Finally, a longer term project will address the conjecture that for irrational $\alpha$ the spectrums of all of these operators are Cantor sets (regardless of the value of $\kappa$). Preliminary numerical results supporting this conjecture are presented in Figures~\ref{fig:allwidths}--\ref{fig:UkHzoom}. Figure~\ref{fig:allwidths} shows the combined width, $W$, of all the spectral bands of $\widetilde U_{kH}(\kappa,\alpha,\lambda)$ as a function of $q$ with fixed $\kappa = 1$ and for several values of $\lambda$. In particular, it seems that $W(q)$ for $\lambda = 1$ approximately follows a power law in $q$ with negative exponent, suggesting that the spectrum's measure could be zero for $\lambda = 1$ and irrational $\alpha$. Figure~\ref{fig:UkHzoom} shows the imaginary part of the logarithm of the spectrum of $\widetilde U_{kH}(\kappa,\alpha,\lambda)$ for $\kappa = \lambda = 1$. This way of displaying the data was chosen for easy comparison when zooming in on different parts of the spectrum, and simply amounts to plotting on a line rather than the circles used in Figures~\ref{fig:Harp}--\ref{fig:kRot1theta}. The figure uses $\alpha = p/q = 2584/4181$, which approximates the reciprocal $(\sqrt {5} - 1)/2$ of the Golden Mean $(\sqrt {5} + 1)/2$, and can be obtained from truncating its continued fraction expansion. The columns show how one obtains strikingly similar structures by repeatedly zooming in on the central region of the spectrum.

We briefly outline a project strategy for proving the conjecture of the spectrums being Cantor sets. The proof that the spectrums of almost Mathieu operators for irrational $\alpha$ are Cantor sets is based on the fact that these operators are represented on $\ell^2(\ZZ),$ via the Fourier transform, by tridiagonal (Jacobi) matrices. Therefore, the Fourier transforms of the generalized eigenfunctions that correspond to the points in the spectrums are sequences in $\ell^{\infty}(\ZZ)$ that satisfy a three term recursion relationship that is expressed using $2 \times 2$ transfer matrices. Unfortunately, the kicked operators can not be represented, via the Fourier transform, by tridiagonal matrices or even by banded matrices. However, a powerful result about loop groups can be used to approximate the kicked operators by unitary operators that can be represented on $\ell^2(\ZZ)$ by banded matrices and hence enable the use of transfer matrices. This result is Proposition~$3.5.3$ in the book \textit{Loop~Groups} by Pressley and Segal \cite{pressley86} which states "If $G$ is semisimple, then $L_{pol}G$ is dense in $LG.$ Here $G$ means a semisimple group of matrices, such as the group $SU(q)$ of $q \times q$ unitary matrices with determinant one, $LG$ is the loop group of $G$ that consists of continuous functions from $\TT$ into $G$ under pointwise multiplication, and $L_{pol}G$ denotes the subgroup of $LG$ consisting of matrix--valued functions each of whose entries is a trigonometric polynomial. This result can be understood as a version of the Weierstrass approximation theorem that asserts that every function in $C(\TT)$ can be uniformly approximated by trigonometric polynomials. This result was used by Lawton in \cite{lawton04} to derive a result about a class of filters that can be used to construct orthonormal bases of wavelets having compact support, and used in combination with operator splitting methods by Oswald in \cite{oswald08} to show that the error of approximating of an element $a \in LG$ by an element $b \in L_{pol}G$ decreases with it smoothness of $a$ and the degree of $b$ in roughly the same manner as for the Weierstrass's theorem. In particular, since the loop group elements related to kicked operators are real analytic functions from $\TT$ into $SU(q),$ the approximation errors decay exponentially with the degree of the approximating loop group elements in $L_{pol}\, SU(q).$ This degree corresponds to the size of the transfer matrices. We note that loop groups have already found extensive applications in physics ranging from the Toda equation \cite{nirov07} to M-theory \cite{adams02}.
\section*{Acknowledgments}
\label{acknowledgments}
W.~L. and A.~S.~M. wish to thank Professor Florin--Petre Boca for his valuable help. A.~S.~M. is supported by a Villum~Kann~Rasmussen grant. J.~W.~and J.~G.~thank Professor C.--H.~Lai for his support and encouragement. J.~W. acknowledges support from the Defence
Science and Technology Agency (DSTA) of Singapore under
agreement POD0613356. J.~G. is supported by
start--up funding (WBS grant No. R-144-050-193-101 and
No. R-144-050-193-133) and the NUS ``YIA'' funding
(WBS grant No. R-144-000-195-123), National University
of Singapore.
\appendix
\section{Physical considerations and experimental realizations}
\label{appendix1}
This appendix reviews some of the physical considerations and difficulties in experimentally realizing the operators considered in Sections~\ref{derivations} and \ref{algorithms}.

The first family of operators treated above, $H(\alpha, \lambda, \theta)$ in Equation~\ref{eq:H1}, arises from the Harper model, which is a tight-binding model describing electrons on a two dimensional square lattice with lattice spacing $a$, subjected to a uniform magnetic field with magnitude $B$ and direction perpendicular to the lattice; see \cite{hofstadter76} for a detailed description. The two-dimensional problem can be separated, yielding Equation~\ref{eq:H1} in one direction, and plane-wave behavior perpendicular hereto. The parameter $\theta$ is proportional to the momentum of this plane-wave, and the parameter $\alpha$ is given by
\begin{eqnarray}
\label{alpha}
\alpha &=& \frac{e a^2}{h}B,
\end{eqnarray}
where $e$ is the elementary charge and $h$ is Planck's constant.
For parameters describing realistic materials, $a \approx 2$\AA, so to have $\alpha \approx 1$ would require $B \approx 10^5$T - orders of magnitude larger than what can be obtained in present laboratories. Nevertheless, other physical situations provide realizations of the Harper model with non--vanishing $\alpha$, yielding the first experimental indication of a Hofstadter's butterfly spectrum in \cite{Schlosser96}.

Following the rules of quantum mechanics, the Unitary Harper operators $U_{H}(\kappa,\alpha,\lambda,\theta)$ defined in Equation~\ref{eq:Uh1} can be seen as the time propagator for a system with a time-independent Hamiltonian given by $H(\alpha, \lambda, \theta)$, where the time interval propagated is proportional to $\kappa$. Thus, since the Harper operators have been experimentally realized, so have these unitary operators.

In contrast, the unitary kicked Harper operators $U_{kH}(\kappa,\alpha,\lambda,\theta)$ defined in Equation~\ref{eq:Ukh1} have not yet been experimentally realized, even though they play an important role in the field of quantum chaos. The possibility of experimentally realizing operators with the same spectrum as the kicked Harper operators is what currently makes the on--resonance double kicked rotor particularly interesting from a physical point of view.

The unitary on--resonance double kicked operators defined in Equation~\ref{eq:Uordkr1},
\begin{equation}
U_{ordkr}(\kappa,\alpha,\lambda,\theta) =
        \exp \Big[ -i 2 \kappa \, \cos(2\pi x) \, \Big] \,
        T(\alpha) \, \exp \Big[ -i 2 \kappa \, \lambda \, \cos\big(2\pi(x+\theta)\big) \, \Big] \, T(\alpha)^{-1}, \nonumber
\end{equation}
were proposed to be experimentally realized in a cold-atom setup \cite{wanggong08}. Apart from the choice of experimental parameters, similar experiments have been performed earlier, see e.g. \cite{jones04}. Specifically, the proposal considered a long quasi--one dimensional gas, consisting of cold atoms or a condensate, subjected to a time--periodic sequence of potential kicks caused by a standing wave generated by two counter--propagating pulsed laser beams. The kicking potential is sinusoidal with spatial period equal to half the lasers wavelength, and $x$ signifies position within such a spatial period. The time dependent Hamiltonian is thus composed of brief time intervals where the dynamics are dominated by the kicking potential, and in between these kicks, time intervals where the dynamics are given by free (kinetic energy only) propagations. This Hamiltonian commutes at all times with translation by half a wavelength, so assuming that the initial state also has this periodicity, the state at all later times will have it as well. We can thus restrict our attention to one period $x \in [0,1)$. We shall have more to say of this periodicity assumption below.

The operator $U_{ordkr}(\kappa,\alpha,\lambda,\theta)$ is the propagator for one time--period of kicks and free propagation, say from time $t_j$ to time $t_{j+1}$. Reading the terms in $U_{ordkr}(\kappa,\alpha,\lambda,\theta)$ from right to left, the $j$'th time--period starts at time $t = t_j$ and contains a free propagation for a time interval $\tau \propto \alpha$ followed by a first kick at time $t = t_j + \tau$. Hereafter, a second free propagation follows for a time interval $T_d - \tau$, and a second kick at time $T_d$ ends the propagation at time $t_{j+1} = t_j + T_d.$ The delay time $T_d$ will be chosen to be the so--called resonance-- or recurrence--time explained below. We turn now to look at the individual terms in $U_{ordkr}(\kappa,\alpha,\lambda,\theta)$ in more detail.

The kicks are the cause of the terms $\exp \left[ -i 2 \kappa \, \cos(2\pi x) \right]$ and $\exp \big[-i 2 \kappa \, \lambda \, \cos\big(2\pi(x+\theta)\big) \, \big]$, where $\theta$ is the spatial separation between the maxima of the two kicking potentials. This separation can be experimentally realized by phase shifting the laser pulses.

The two terms $T(\alpha)$ and $T(\alpha)^{-1}$ are due to free propagation between the kicks. Whereas $T(\alpha)^{-1}$ appears directly so in the propagator, $T(\alpha)$ actually arises from the free propagation for a time interval $T_d - \tau$, yielding $\exp\left[ \frac{i T_d}{2\pi T_R} \frac{d^2}{dx^2}\right] \, \exp \left[\frac{-i\alpha}{4\pi} \frac{d^2}{dx^2} \right].$ Therefore, choosing the delay time equal to the resonance time, $T_d = T_R$, the first term in the operator becomes $\exp\left[\frac{i}{2\pi} \frac{d^2}{dx^2} \right]$, which can be ignored since it equals $1$ when applied to any basis function $\xi_n(x)$. In passing, we notice that the existence of such a time is due to the free propagation having eigen--energies, which are all an integer multiple of a certain energy. We also note that the experimentally realizable ranges for the parameters have been estimated to be $\alpha \in [0.005, 2],$ $\kappa \in [0.1,100]$ and $\kappa \lambda \in [0.1,100]$, whereas $\theta$ can take on all values \cite{wang08}.

Returning to the assumption that the initial state is periodic in $x$, we now consider a somewhat more general initial state, namely a Bloch wave $\Psi(x) = \psi(x) \exp(i 2 \pi \, \theta_B \, x)$, where the wavefunction $\Psi(x)$ on the entire real axis is a product of a function $\psi(x)$ having period $1$, and a complex exponential with quasimomentum proportional to $\theta_B$. This quasimomentum in the wavefunction would correspond to a real momentum of the entire gas or condensate cloud in the experimental setup. As above, the translational symmetry of the Hamiltonian ensures that $\theta_B$ does not change in time. Again, we can restrict our considerations to functions on the interval $[0,1)$ with the basis $\left\{ \xi_n(x) \right\}$, provided we make the substitution $d/dx \, \rightarrow \, d/dx + i 2 \pi \, \theta_B$. Making these substitutions, and again choosing $T_d = T_R$, we see that
\begin{eqnarray}
\label{eq:Udkrquasimom}
U_{ordkr}(\kappa,\alpha,\lambda,\theta) &\rightarrow&
\exp\Big[-i 2 \pi \, \theta_{B}^{\, 2} \Big] \, U_{ordkr}\Big(\kappa, \alpha, \lambda, \theta + \theta_B(\alpha-2) \Big) \, \exp\left[ -2 \, \theta_B \, \frac{d}{dx} \right]. \nonumber
\end{eqnarray}
Thus, the introduction of $\theta_B$ gives rise to three changes. First, an extra numerical term is introduced, which has the effect of moving the eigenvalues on the circle. Second, in $U_{ordkr}(\kappa, \alpha, \lambda, \theta)$, the parameter $\theta \rightarrow \theta + \theta_B(\alpha-2)$. Third, an extra operator $\exp\left[ -2 \, \theta_B \, d/dx \right]$ is right--multiplied on $U_{ordkr}$. With this last extra term, the resulting operator no longer belongs to the rotation $C^*$--algebra, and its spectrum can be very different from the original operator \footnote{This pronounced difference is apparent when comparing Figure~1a with Figure~3c in \cite{wangmouritzengong08} for the value $\hbar/\pi = \alpha/4 = 1/2$. Figure~1a displays the imaginary part of the logarithm of the spectrum of $U_{ordkr}(\kappa, \alpha, \lambda, \theta)$, while Figure~3c displays the imaginary part of the logarithm of the spectrum of $U_{ordkr}(\kappa, \alpha, \lambda, \theta) \, \exp\left( -\frac{1}{2} \frac{d}{dx} \right)$, both using the parameters $\kappa = \lambda = 1$ and $\theta = 0$. To see that the latter operator equals the operator used to calculate the spectrums in Figure~3c, which uses the anti--resonance condition $T_d = T_R/2$, observe that $\exp\left[\pm \frac{i}{4\pi} \frac{d^2}{dx^2} \right]$ and $\exp\left[ \pm \frac{1}{2} \frac{d}{dx} \right]$ give the same result when applied to any basis function $\xi_n(x).$ Thus, for the value $\hbar/\pi = 1/2$, the operator used to find Figure~3c corresponds to $\exp(i \pi/8)$ times the operator on the right hand side of Equation~\ref{eq:Udkrquasimom} with $\theta_B = 1/4$. The only effect of the factor $\exp(i \pi/8)$ is to shift the spectrum, so the spectral differences are easily seen.}. Notably, although the $\theta$ in the kicked Harper operator, Equation~\ref{eq:Ukh1}, can be thought of as coming from introducing a quasimomentum by $d/dx \rightarrow d/dx + i 2 \pi \, \theta / \alpha$, this procedure has little connection to the quasimomentum in the experiment proposed in \cite{wanggong08}.

An interesting question for future research is thus how this assumption of $\theta_B = 0$ influences the experimental realizability of $U_{ordkr}(\kappa,\alpha,\lambda,\theta)$, and if it can be relaxed. In a real experiment, the cold atoms or condensate is trapped in a weak external potential, making the initial wavefunction a wide gaussian rather than a Bloch wave, further complicating the situation. That being said, in the usual experimental situation, the wavefunction's finite width gives negligible contributions to the outcome of many presently performed measurements.

Finally, it should be noted that if one were instead to use a ring-shaped trap, the quasimomentum is exactly zero, and realizing the operator $U_{ordkr}(\kappa,\alpha,\lambda,\theta)$ would not be troubled by the issues raised above. Rather than using kicks from a laser, one must then use potential kicks that are linear in space, which would yield exactly the cosines in the angular coordinate of the ring, and the $\theta$--parameter would be the angle between the directions of the spatially linear kicks.
\section{Spectral Theory and $C^*$--Algebras}
\label{appendix2}
This appendix summarizes results in \cite{kato66} and \cite{murphy90} that explain the relationship between the structure of an algebra and the spectrums of elements in it. It also derives auxiliary results used in Section \ref{derivations} and Appendix~\ref{appendix3}.
\begin{definition}
\label{algebra}
An algebra is a complex vector space $\cA$ equipped with a bilinear
map
$\cA^2 \rightarrow \cA, (A,B) \rightarrow AB,$
$($called multiplication$)$ that satisfies $A(BC) = (AB)C.$ An algebra $\cA$
is called abelian if $AB = BA,$ unital if it has an $($unit or identity$)$
element $I_{\cA}$ with $I_{\cA}A = AI_{\cA} = A$ and
$A \in \cA$ is called invertible if
there exists an element $B$ such that $AB = BA = I_{\cA}.$
Then $B$ is unique, is called the inverse of $A,$ and is denoted by $A^{-1}.$
The set of all invertible elements is denoted by ${\textnormal{Inv}}(\cA),$
and for every element $A \in \cA$ its spectrum is defined by
\begin{equation}
\label{specdef}
    \sigma_{\cA}(A) = \big\{ \, \mu \in \CC \, : \, \mu \, I_{\cA} - A \, \not \in \, \textnormal{Inv}(\cA) \, \big\},
\end{equation}
and its spectral radius is defined by
$r(A) = \max \{ \, |\mu| \, : \, \mu \in \sigma(A) \, \}.$
A subspace $J \subseteq \cA$ is called a subalgebra if
$
    A, B \in J \implies AB \in J.
$
We define the center of an algebra $\cA$ by
$$\cZ(\cA) = \big\{ \, A \in \cA \, : \, AB = BA \ \textnormal{for every} \, B \in \cA \, \big\}.$$
Clearly $\cZ(\cA)$ is a subalgebra of $\cA.$
A subset $J \subset \cA$ is an ideal of an algebra $\cA$ if
$
    A \in \cA, B \in J \ \implies AB, BA \in J.
$
The ideals $\{0\}$ and $\cA$ are called trivial and an algebra with only
trivial ideals is called simple.
A homomorphism from an algebra $\cA$ to an algebra $\cB$ is a linear
map $\phi \, : \, \cA \rightarrow \cB$ that satisfies
$
    \phi(AB) = \phi(A)\phi(B).
$
$\phi$ is called unital if $\cA$ and $\cB$ are unital and $\phi(I_{\cA}) = I_{\cB}.$
The kernel
$\textnormal{ker}(\phi) = \{ \, A \in \cA \, : \, \phi(A) = 0 \, \}$
is an ideal and $\phi$ is injective if and only if $\textnormal{ker}(\phi) = \{0\}.$
$\phi$ is called an isomorphism if it is bijective, and then we say that $\cA$ and
$\cB$ are isomorphic and write $\cA \cong \cB.$ If $J \subseteq \cA$
is an ideal then the set of cosets $\cA \, / \, J$ is an algebra under the multiplication
$(A+J)(B+J) = AB + J$ and the map $\rho \, : \, \cA \rightarrow \cA \, / \, J$
defined by $\rho(A) = A + J$ is a homomorphism and $\textnormal{ker}(\rho) = J.$
Every homomorphism $\phi \, : \, \cA \rightarrow \cB$ induces an isomorphism
$\widetilde {\phi} \, : \, \cA \, / \, \textnormal{ker}(\phi) \rightarrow \phi(\cA).$
\end{definition}
\begin{lemma}
\label{spec1}
If $\cA$ is a unital algebra and $\cB$ is a subalgebra that contains $I_{\cA},$ then
$\sigma_{\cA}(B) \subseteq \sigma_{\cB}(B)$ for every $B \in \cB.$

\end{lemma}
{\it Proof.} If $\mu \not\in \sigma_{\cB}(B)$ then
$\mu \, I_{\cA} - B \, = \, \mu \, I_{\cB} - B \, \in \, \textnormal{Inv}(\cA)$ hence
$\mu \not\in \sigma_{\cA}(B).$ $\qquad \blacksquare$
\begin{definition}
\label{staralgebra}
An involution on an algebra $\cA$ is a map $A \rightarrow A^*$ that satisfies
$(A+B)^* = A^* + B^*, \ (AB)^* = B^*A^*, \ A^{**} = A, \ (cA)^* = \overline{c} \, A^*$
and $I_{\cA}^{*} = I_{\cA}$ whenever $\cA$ is unital. Then $\cA$ is called a $*$-algebra. If $\cA$ and $\cB$ are $*$-algebras
then a homomorphism $\phi \, : \, \cA \rightarrow \cB$ is called a $*$-homomorphism if it satisfies $\phi(A^*) = \big(\phi(A)\big)^*.$
An element $A$ of a $*$-algebra $\cA$ is called normal if $AA^* = A^*A,$ self-adjoint if $A^* = A,$ and unitary if $A \in \textnormal{Inv}(\cA)$
and $A^* = A^{-1}.$
\end{definition}
\begin{definition}
\label{normalgebra}
A normed algebra is an algebra together with a vector norm
$||\cdot||$ that satisfies
$
    ||AB|| \leq ||A|| \, ||B||.
$
In this paper we only consider continuous homomorphisms between normed algebras.
A homomorphism that preserves norms is called an isometry.
A Banach algebra is a complete normed algebra. If $\cA$ is a unital Banach algebra and $A \in \cA$ then $\cC(A)$ denotes the $($necessarily abelian$)$ norm completion of the set of all polynomials in $I_{\cA}$ and $A$ and is called the Banach algebra generated by $A.$
\end{definition}
\begin{definition}
\label{Cstaralgebra}
A Banach $*$--algebra is a Banach algebra with an involution $*$ that satisfies $||A^*|| = ||A||.$ A $C^*$--algebra is a Banach $*$--algebra that satisfies $||A^*A|| = ||A||^2.$ If $\cA$ is a unital $C^*$--algebra and $S \subseteq \cA$ we define the $C^*$--algebra generated by $S$ to be the smallest $C^*$--subalgebra of $\cA$ that contains $I_{\cA}$ and $S.$ If $\cA$ is a unital $C^*$--algebra and $A \in \cA$ we let $\cC^*(A)$ denote the $C^*$--algebra generated by $\{A\}.$ Clearly, $\cC^*(A)$ is the $($necessarily abelian$)$ norm completion of the set of polynomials in $I_{\cA}, A$ and $A^*.$
If $\cA$ and $\cB$ are unital $C^*$--algebras, then $\textnormal{Hom}(\cA,\cB)$ will denote the Banach space, i.e. the complete normed vector space, of $*$--homomorphisms from $\cA$ to $\cB$ equipped with the operator norm topology. If $\cA$ and $\cB$ are unital then we require that homomorphisms map $I_{\cA}$ to $I_{\cB}.$
\end{definition}
{\bf Examples.} $C(\TT)$, $L^{\infty}(\TT)$, and $\ell^{\infty}(\ZZ)$
are Banach spaces under the norm $||\cdot||_{\infty},$
$L^{1}(\TT)$ and $\ell^1(\ZZ)$ are Banach spaces under the norm
$||\cdot||_{1},$ and $L^2(\TT),$ $\ell^{2}(\ZZ)$ and $\CC^q, \,
q \in \NN$ are Hilbert spaces. For $n \in \ZZ$ we define $\delta_n \, : \, \ZZ \rightarrow \{\, 0, \, 1 \, \}$
by $\delta_n(m) = 1$ for $m = n$ and $\delta_n(m) = 0$ for $m \neq n.$
The Banach spaces $L^{1}(\TT)$ and $\ell^1(\ZZ)$ are Banach $*$--algebras whose multiplication is convolution
and whose involution is defined by $f^{*}(x) = \overline {f(-x)}.$ Only $\ell^1(\ZZ)$ is unital and neither are $C^*$--algebras. For
instance the function $f = -2\delta_{-1} + \delta_{0} + \delta_{1} \in \ell^1(\ZZ)$ satisfies $||f*f^*||_{1} = 12 < 16 = ||f||_{1}^{2}.$ The Banach spaces $C(\TT)$, $L^{\infty}(\TT)$, and $\ell^{\infty}(\ZZ)$ are abelian $C^*$--algebras whose multiplication is pointwise multiplication and whose involution is complex conjugation.
The Hardy subspace $H^{\infty}(\TT) \subset L^{\infty}(\TT),$ consisting of functions $f$ such that $\langle \xi_n\, | \, f \rangle = 0$ whenever $n < 0,$ is a Banach subalgebra of $L^{\infty}(\TT),$ but not a $*$--subalgebra.
The algebra $\cB(\HH)$ of bounded operators on a Hilbert space $\HH$ whose multiplication is composition, whose involution is the adjoint map and whose norm is the operator norm, is a unital $C^*$--algebra. It is abelian if and only if dim$(\HH) = 1.$ It is simple if and only if $q = \dim(\HH) < \infty$ and then it is isomorphic to the algebra $\cM_q$ of $q$ by $q$ matrices. The set $\cK(\HH)$ of compact operators is an ideal and a simple $C^*$--algebra (\cite{murphy90}, Example 3.2.2) and every automorphism of $\cK(\HH)$ is an inner automorphism, viz.~of the form $A \rightarrow UAU^{-1}$ for some $U \in \cB_u(\HH)$, see (\cite{murphy90}, Theorem~$2.4.8$) and (\cite{davidson96}, Lemma~$\textnormal{V}.6.1$). Furthermore, $\cK(\HH)$ is unital if and only if $q = \dim(\HH) < \infty$ and then $\cK(\HH)$ is isomorphic to $\cM_q.$ The $C^*$--algebra $\cC(\HH) = \cB(\HH) \, / \, \cK(\HH)$ is called the Calkin algebra and it is simple. If $\dim(\HH) = \infty$ then this algebra has unitary elements that have no logarithm (\cite{murphy90}, Example 1.4.4).
\begin{lemma}
\label{spec2}
If $\cA$ is a unital Banach algebra and $A \in \cA,$ then $\sigma_{\cA}(A)$ is a nonempty subset of $\DD\big(||A||\big).$
\end{lemma}
{\it Proof.} Lemma~$1.2.4$ and Theorem~$1.2.5$ in \cite{murphy90}. $\qquad \blacksquare$
\begin{lemma}
\label{beurling}
(Beurling) If $\cA$ is a unital Banach algebra, then the spectral radius of every element $A \in \cA$ satisfies
\begin{equation}
\label{eq:beurling}
r(A) \, = \, \inf_{n \geq 1} ||A^n||^{\frac{1}{n}} \, = \, \lim_{n \rightarrow \infty} ||A^n||^{\frac{1}{n}}.
\end{equation}
\end{lemma}
{\it Proof.} Theorem~$1.2.7$ in \cite{murphy90}. $\qquad \blacksquare$
\begin{lemma}
\label{spec3}
If $\cA$ is a unital Banach algebra and $\cB \subseteq \cA$ is a Banach subalgebra that contains $I_{\cA},$ then for every ${B \in \cB},$ $\sigma_{\cA}(B) \subseteq \sigma_{\cB}(B)$ and $\partial \sigma_{\cB}(B) \subseteq \partial \sigma_{\cA}(B)$ $($where the boundary $\partial X = \hbox{closure}(X) \cap \hbox{closure}(\CC \backslash X)$ for $X \subset \CC${}$)$. Furthermore,
if $\sigma_{\cA}(B)$ has no holes $($viz. its complement is topologically connected$)$ then $\sigma_{\cA}(B) = \sigma_{\cB}(B).$
\end{lemma}
{\it Proof.} Lemma~\ref{spec1} implies that $\sigma_{\cA}(B) \subseteq \sigma_{\cB}(B).$
Theorem~$1.2.8$ in \cite{murphy90} implies the remaining assertions. $\qquad \blacksquare$
\\ \\
{\bf Example.} If $B = z \in \cB = H^{\infty}(\TT) \subset \cA = L^{\infty}(\TT),$
then $\sigma_{\cA}(B) = \TT_c \subseteq \DD(1) = \sigma_{\cB}(B)$ and $\partial \DD(1) = \TT_c = \partial \TT_c.$
\begin{definition}
\label{spectrumalgebra}
$($Gelfand$)$ The spectrum $($as defined on page~15 in \cite{murphy90}\, $)$ of an abelian Banach algebra $\cA$ is the set $\Omega(\cA)$ of nonzero continuous homomorphisms from $\cA$ to $\CC$ equipped with the weak$^*$ topology $($explained in the Appendix in \cite{murphy90}\, $)$.
The Gelfand transform of $A \in \cA$ $($also defined on page~15 in \cite{murphy90}\, $)$ is the function
$\widehat A \, : \, \Omega(\cA) \rightarrow \CC$ defined by
$
    \widehat A (\omega) = \omega(A), \ \omega \in \Omega.
$
\end{definition}
\begin{lemma}
\label{gelfand1}
$($Gelfand$)$ If $\cA$ is an abelian unital Banach algebra, then
\begin{enumerate}
\item $\Omega(\cA)$ is nonempty and compact,
\item $\widehat A \in C\big(\Omega(\cA)\big)$ and
        $\sigma(A) = \widehat {A}\big(\Omega(\cA)\big)$
            for every $A \in \cA,$
\item the map $A \rightarrow \widehat A$ is a norm--decreasing homomorphism from
    $\cA$ into $C\big(\Omega(\cA)\big).$
\end{enumerate}
Furthermore, if $A \in \cA$ and if $\cA$ is generated by $I_{\cA}$ and $A,$ then $\cA$ is abelian and the map $\widehat {A} \, : \, \Omega(\cA) \rightarrow \sigma(A)$ is a homeomorphism $($i.e.~a continuous bijective map whose inverse is also continuous$).$
\end{lemma}
{\it Proof.} Theorems 1.3.3, 1.3.5, 1.3.6 and 1.3.7 in \cite{murphy90}. $\qquad \blacksquare$
\\ \\
{\bf Example.} The Gelfand transform for $\ell^1(\ZZ)$ is the
Fourier transform
$\cF \, : \, \ell^1(\ZZ) \rightarrow C(\TT),$
which is defined by
$\cF(f) = \sum_{n \in \ZZ} \, f(n) \, \xi_n.$
It is injective. However, since the function
$f \, = \, -2\delta_{-1} + \delta_{0} + \delta_{1} \, \in \, \ell^1(\ZZ)$
satisfies
${||\cF(f)||_{\infty} = \sqrt {35}/2 < 4 = ||f||_{1}},$
it is not an isometry.
$\cF\big(\ell^1(\ZZ)\big) \subset C(\TT)$ consists of continuous functions whose Fourier series are absolutely convergent.
Wiener \cite{wiener32} first proved that if $h \in \cF\big(\ell^1(\ZZ)\big)$ and $h$ never vanishes then $1/h \in \cF\big(\ell^1(\ZZ)\big).$
\begin{lemma}
\label{gelfand2}
$($Gelfand$)$ If $\cA$ is a non--zero abelian unital $C^*$--algebra, then the map $A \rightarrow \widehat A$ is an isometric \hbox{$*$--isomorphism} from $\cA$ onto $C\big(\Omega(\cA)\big).$ If $\cA$ is a unital $C^*$--algebra $($not necessarily abelian$)$ then
$r(A) = ||A||$ for every normal element $A.$ If $\cA$ is a unital $C^*$--algebra $($not necessarily abelian$)$ and $\cB \subseteq \cA$ is a $C^*$--subalgebra that contains $I_{\cA},$ then $\sigma_{\cA}(B) = \sigma_{\cB}(B)$ for every $B \in \cB.$
\end{lemma}
{\it Proof.} Theorem~$2.1.10$ in \cite{murphy90} implies the first assertion directly and implies the second assertion by applying it to $\cC^*(A).$ Theorem~$2.1.11$ in \cite{murphy90} is the third assertion. $\qquad \blacksquare$
\\ \\
In view of Lemma~\ref{gelfand2}, we will denote the spectrum of an element $A$ of a unital $C^*$--algebra simply by $\sigma(A)$ since the spectrum is independent of the algebra with respect to which it is defined by Equation~\ref{specdef}. Furthermore, we observe that if $\cA$ is a unital $C^*$--algebra and if
$A \in \cA$ is normal, then every
$\phi \in \Omega\big(\cC(A)\big)$
can be extended to
$\widetilde {\phi} \in \Omega\big(\cC^*(A)\big)$
by setting
$\widetilde {\phi}(A^*) = \overline {\phi(A)}.$
The map
$\phi \rightarrow \widetilde {\phi}$
is a bijection from $\Omega\big(\cC(A)\big)$ onto $\Omega\big(\cC^*(A)\big).$
Therefore, we may identify $\Omega\big(\cC(A)\big)$ with $\Omega\big(\cC^*(A)\big)$
and then Lemma~\ref{gelfand1} implies that $\widehat A$ may be identified with a homeomorphism of $\Omega\big(\cC^*(A)\big)$ onto $\sigma(A).$ Then we may identify $A$ with the restriction of the identity map $z \, : \, \CC \rightarrow \CC$ to $\sigma(A)$ and identify $A^*$ with the restriction of $\overline z$ to $\sigma(A).$ Therefore, we may identify $\cC^*(A)$ with the $C^*$--subalgebra $\cC^*(z)$ of $C\big(\sigma(A)\big)$ that is generated by the functions $1, z$ and $\overline z.$ The Stone-Weierstrass Theorem implies that the algebra of functions on $\sigma(A)$ that are defined by polynomials in $z$ and $\overline z$ is dense in $C\big(\sigma(A)\big)$ and hence that $\cC^*(z) = C\big(\sigma(A)\big).$ Therefore, we may identify
$\cC^*(A)$ with $C\big(\sigma(A)\big).$
\begin{lemma}
\label{specmap}
$($Spectral Mapping Lemma$)$ If $\cA$ is a unital $C^*$--algebra, $A \in \cA$ is normal,
and $f \in C\big(\sigma(A)\big),$ then $f(A) \in \cA$ and $\sigma\big(f(A)\big) = f\big(\sigma(A)\big).$
\end{lemma}
{\it Proof.} As noted above, since $A$ is normal $\cC^*(A)$ is an abelian $C^*$--algebra that can be identified with $C\big(\sigma(A)\big)$ and $A$ can be identified with the restriction of the identity function $z$ to $\sigma(A).$ Therefore, $f(A)$ can be identified with the element $f(z) \in C\big(\sigma(A)\big) \subseteq \cA.$ The second assertion is Theorem~$2.1.14$ in \cite{murphy90}. $\qquad \blacksquare$
\begin{pro}
\label{dist}
If $\cA$ is a unital $C^*$--algebra then $d\big[\sigma(A),\sigma(B)\big] \, \leq \, || \, A - B \, ||$ for normal $A, B \in \cA.$
\end{pro}
{\it Proof.} Let $s = d\big[\sigma(A),\sigma(B)\big]$ and assume to the contrary that $s > || \, A - B \, ||.$ Without loss of generality (by swapping $A$ and $B$) we can assume that there exists an $a \in \sigma(A)$ such that
$s = \min \{ \, |a-b| \, : \, b \in \sigma(B) \, \}.$
We observe that the spectrum of $B$ is disjoint from the open disc of radius $s$ centered at $a.$ Therefore, the operator $aI_{\cA}-B$ is invertible and normal and its spectrum is disjoint from the open disc of radius $s$ centered at $0.$ Furthermore, we observe that the map $z \rightarrow z^{-1}$ maps the union of $\{\infty\}$ with the complement of this open disc onto the closed disc $\DD(1/s).$ Therefore, we can apply Lemma~\ref{specmap} to obtain
$\sigma\big( (aI_{\cA}-B)^{-1}\big) \subseteq \DD(1/s).$
Lemma~\ref{gelfand2} implies that $||(aI_{\cA} - B)^{-1}|| = r\big((aI_{\cA} - B)^{-1}\big)$ and hence
$||(aI_{\cA} - B)^{-1}|| \leq 1/s.$
Using the multiplicative property of the norm given in Definition~\ref{normalgebra} we obtain
$$\big|\big|(aI_{\cA}-A)(aI_{\cA}-B)^{-1} - I_{\cA}\big|\big| \, = \, \big|\big|\, (aI_{\cA}-B)^{-1}\, \big((aI_{\cA}-A) - (aI_{\cA}-B)\, \big)\, \big|\big|
\, \leq \, \big|\big|(aI_{\cA}-B)^{-1}\big|\big| \, ||A-B|| \, < \, 1.$$
A standard result (\cite{murphy90}, Theorem~$1.2.2$) implies that
$(aI_{\cA}-A)(aI_{\cA}-B)^{-1} \in Inv(\cA)$
and hence $aI_{\cA}-A \in Inv(\cA).$ This contradicts the
fact that $a \in \sigma(A)$ and concludes the proof. $\qquad \blacksquare$
\begin{pro}
\label{separate}
Let $\cA$ and $\cB$ be unital $C^*$--algebras, let $X$ be a compact topological space, and let
$\rho \, : \, X \rightarrow \textnormal{Hom}(\cA,\cB)$ be a continuous map such that for every $A \in \cA$ there exists an $x \in X$ such that $\rho(x)(A) \neq 0$ $($then we say that the subset $\rho(X) \subseteq \textnormal{Hom}(\cA,\cB)$ separates points$)$. Then for every normal $A \in \cA,$
\begin{equation}
\label{union}
    \sigma(A) = \bigcup_{x \in X} \sigma\big(\rho(x)(A)\big).
\end{equation}
\end{pro}
{\it Proof.} Let $A \in \cA$ be normal. If $\mu \in \CC$ is not in the spectrum of $A$ then $\mu I_{\cA} - A$ has an inverse and hence for every $x \in X,$ $\mu I_{\cB} - \rho(x)(A)$ has the inverse
$\rho(x)\left[(\mu I_{\cA} - A)^{-1}\right]$
and therefore $\mu \not \in \sigma\big(\rho(x)(A)\big).$ This proves that $\sigma\big(\rho(x)(A)\big) \subseteq \sigma(A)$ for every $x \in X.$
We will now show that the opposite inclusion holds. We first observe that since $A$ is normal,
Lemma~\ref{gelfand2} implies that without loss of generality we may assume that $\cA = \cC^*(A).$ For every $x \in X$ we denote $\rho(x)(A)$ by $A_{x}$ and the subalgebra $\rho(x)(\cA) \subseteq \cB$ by $\cA_{x}.$ Clearly, $\cA_x = \cC^*(A_x).$
We observe that each homomorphism $\rho(x) \, : \, \cA \rightarrow \cA_x$ induces a continuous
map $\widetilde {\rho(x)} \, : \, \Omega(\cA_x) \rightarrow \Omega(\cA)$ defined by
$\widetilde {\rho(x)}(\phi) = \phi \circ \rho(x), \ \phi \in \Omega(\cA_x).$
The discussion following Lemma~\ref{gelfand2} shows that we may identify $\Omega(\cA)$ with $\sigma(A),$ $\cA$ with $C\big(\sigma(A)\big),$ and each $\cA_{x}$ with $C\big(\sigma(A_x)\big).$
With these identifications $\widetilde {\rho(x)} \, : \, \sigma(A_x) \rightarrow \sigma(A)$ is the
inclusion map. Furthermore, Lemma~\ref{specmap} implies that for each $f \in C\big(\sigma(A)\big)$ and each $x \in X$ we may identify $\rho(x)(f)$ with the restriction of $f$ to the subset $\sigma(A_x)$ of $\sigma(A).$ Let $Y$ denote the closure of $\bigcup_{x \in X} \sigma(A_x).$ Clearly $Y \subseteq \sigma(A).$ We now show that $\rho(X)$ separates the points in $C\big(\sigma(A)\big)$ if and only if $Y = \sigma(A).$ If $Y \neq \sigma(A)$ then there exists a nonzero function $f \in C\big(\sigma(A)\big)$ whose restriction $f|_Y = 0.$ Then for every $x \in X$ the function $\rho(x)(f) = f|_{\sigma(A_x)} = 0$ and hence $\rho(X)$ does not separate the points in $C\big(\sigma(A)\big).$ To prove the converse assume that $Y = \sigma(A)$ and that $f$ is a nonzero function in $C\big(\sigma(A)\big).$ Therefore, $\bigcup_{x \in X} \sigma(A_x)$ is a dense subset of $\sigma(A)$ and hence the restriction of $f$ to $\bigcup_{x \in X} \sigma(A_x)$ is not equal to $0.$ Therefore, there exists an $x \in X$ such that $\rho(x)(f) \neq 0$ and hence $\rho(X)$ separates points in $C\big(\sigma(A)\big).$ We conclude the proof by proving that $Y = \bigcup_{x \in X} \sigma(A_x).$ It suffices to show that $\bigcup_{x \in X} \sigma(A_x)$ is closed. Let $p \in Y$ and let $p_i \in \bigcup_{x \in X} \sigma(A_x), \ i \in \NN$ be a sequence that converges to $p.$ It suffices to show that $p \in \bigcup_{x \in X} \sigma(A_x).$ For every $i \in \NN$ there exists an $x_i \in X$ such that $p_i \in \sigma(A_{x_i}).$ Since $X$ is compact, we may assume without loss of generality (by considering a subsequence of $x_i$) that there exists a $y \in X$ such that $x_i$ converges to $y.$
Proposition~\ref{dist} implies that the map $x \rightarrow \sigma(A_x)$ is continuous. Therefore, the sequence $\sigma(A_{x_i})$ of points in the
hyperspace $\cH\big(\sigma(A)\big)$ converges to $\sigma(A_y).$ Since $p_i \in \sigma(A_{x_i})$ and
$d\big[\sigma(A_{x_i}),\sigma(A_y)\big] \rightarrow 0,$ it follows that there exists a sequence $y_i \in \sigma(A_y)$ such that $|p_i-y_i| \rightarrow 0.$ Therefore, $|y_i - p| \rightarrow 0$ and hence $p = y \in \sigma(A_y).$ This concludes the proof. $\qquad \blacksquare$
\begin{definition}
\label{def:representation}
A representation of a $C^*$--algebra $\cA$ on a Hilbert space $\HH$ is a $*$--homomorphism $\rho \in \textnormal{Hom}\big(\cA,\cB(\HH)\big).$ A representation
$\rho$ is faithful if the kernel of $\rho$ equals $0$ $($the zero subspace of $\HH${}$).$ A representation $\rho$ is irreducible if $0$ and $\HH$ are the only closed subspaces of $\HH$ that are invariant under $\rho(A)$ for every $A \in \cA.$
\end{definition}
\begin{lemma}
\label{gelfandnaimark}
Every $C^*$--algebra is isomorphic to a $C^*$--subalgebra of $\cB(\HH)$ for some Hilbert space $\HH,$ or equivalently, if it admits a faithful representation.
\end{lemma}
{\it Proof.} This is Theorem~$3.4.1$ in \cite{murphy90}. Such a faithful representation is called a Gelfand-Naimark-Segal representation. $\qquad \blacksquare$
\section{Rotation $C^*$--Algebras and Their Representations}
\label{appendix3}
This appendix summarizes results in \cite{boca01} and \cite{davidson96} that explain the structure of rotation $C^*$--algebras. It also derives auxiliary results used in Section~\ref{derivations}.
\begin{definition}
\label{rotdef}
A rotation $C^*$--algebra is a unital $C^*$--algebra $\cA$ that is generated by unitary elements $U$ and $V$ that satisfy the commutation relation
\begin{equation}
\label{commUV}
    UV = \exp(i2\pi \alpha)VU
\end{equation}
for some $\alpha \in [0,1).$ The pair $(U,V)$ is said to be a frame with parameter $\alpha$ for $\cA.$
\end{definition}
{\bf Example 1.} Let $q \in \NN$ and define $\omega = \exp(i2\pi /q)$ and matrices $F, C, D \in \cM_q$ by
\begin{equation}
\label{FCD}
F = \frac{1}{\sqrt{q}}\left[ \begin{array}{cccccc}
        1 & 1 & 1 & \cdots & 1 & 1 \\
        1 & \omega & \omega^2 & \omega^3 & \cdots & \omega^{-1} \\
        1 & \omega^2 & \omega^4 & \omega^6 & \cdots & \omega^{-2} \\
        \vdots & \vdots & \vdots & \vdots & \ddots & \vdots \\
        1 & \omega^{-1} & \omega^{-2} & \omega^{-3} & \cdots & \omega
      \end{array} \right], \qquad
C =
\left[ \begin{array}{cccccc}
0 & 1 & 0 & \cdots & 0 & 0 \\
0 & 0 & 1 & \cdots & 0 & 0 \\
\vdots & \vdots & \vdots & \ddots & \vdots & \vdots \\
0 & 0 & 0 & \cdots & 0 & 1 \\
1 & 0 & 0 & \cdots & 0 & 0
\end{array}\right], \qquad
D =
\left[ \begin{array}{cccccc}
1 & 0 & 0 & \cdots & 0 & 0 \\
0 & \omega & 0 & \cdots & 0 & 0 \\
0 & 0 & \omega^2 & \cdots & 0 & 0 \\
\vdots  &  \vdots & \vdots & \ddots & \vdots & \vdots \\
0  & 0 & 0 & \cdots & 0 & \omega^{q-1}
\end{array}\right].
\end{equation}
Then $(C,D)$ is a frame with parameter $1/q$ for the rotation $C^*$-algebra $\cM_q.$
Furthermore, if $\alpha = p/q \in (0,1)$ with $p$ and $q$ coprime, then $(C^p,D)$ and $(C,D^p)$ are frames with parameter $\alpha$
for $\cM_q.$ The matrix $F$ is called the discrete Fourier transform matrix. We observe that $CF = FD$ hence $C = FDF^{-1}$ is diagonalized by $F.$
This property is extremely useful for developing efficient algorithms.
\\
{\bf Example 2.} Let
    $M \, : \, L^{\infty}(\TT) \rightarrow \cB\big(L^2(\TT)\big)$
and
    $R \, : \, \TT \rightarrow \cB\big(L^2(\TT)\big),$
be defined by Equations \ref{M} and \ref{R}, let $\alpha \in [0,1),$ and let $\cB_\alpha$ denote the $C^*$--subalgebra of $\cB\big(L^2(\TT)\big)$ generated by $R(\alpha)$ and $M(\xi_1).$ Then $\big( R(\alpha), \, M(\xi_1)\big)$ is a frame with parameter $\alpha$ for the rotation $C^*$--algebra $\cB_{\alpha}.$
\begin{pro}
\label{univrot}
For $\alpha \in [0,1)$ there exists a unique $($up to isomorphism$)$ universal rotation $C^*$-algebra $\cA_{\alpha}$ that has the following properties:
\begin{enumerate}
\item there exists a frame $(\widetilde {U}_{\alpha}, \widetilde {V}_{\alpha})$ with
parameter $\alpha$ for $\cA_\alpha,$
\item if $(U,V)$ is a frame with parameter $\alpha$ for a rotation $C^*$--algebra $\cC^*(U,V)$ then
there exists a unique \mbox{$*$--homomorphism} $\pi \, \in \, \textnormal{Hom}\big(\cA_\alpha, \, \cC^*(U,V)\big)$ such that
$\pi(\widetilde {U}_\alpha) = U$ and
$\pi(\widetilde {V}_\alpha) = V.$
\end{enumerate}
Furthermore, this homomorphism is a surjection.
\end{pro}
{\it Proof.} A proof based on the Gelfand-Naimark-Segal construction is given in Chapter VI of \cite{davidson96}. $\qquad \blacksquare$
\\ \\
For $q \in \NN$ define the equivalence relation $\cong_q$ on $\TT_{c}^2$ by
$(z_{1}^{\prime},z_{2}^{\prime}) \cong_q (z_1,z_2)$
if and only if there exists
$(w_1,w_2) \in \TT_{c}^{2}$ such that
$w_{1}^{q} = w_{2}^{q} = 1,$ $z_{1}^{\prime} = w_1\, z_{1},$ and $z_{2}^{\prime} = w_2\, z_{2}.$
\begin{lemma}
\label{irredrep1}
For $\alpha = p/q$ where $q \in \NN$ and $p$ and $q$ are coprime integers and $(z_1,z_2) \in \TT_{c}^{2}$, let
$\pi_{z_1,z_2} \in \textnormal{Hom}(\cA_{\alpha},\cM_q)$
denote the representation of $\cA_{\alpha}$ on $\CC^q$ that satisfies
$\pi_{z_1,z_2}(\widetilde {U}_{\alpha}) = z_1\, C^p$
and
$\pi_{z_1,z_2}(\widetilde {V}_{\alpha}) = z_2\, D.$
We observe that the existence and uniqueness of $\pi_{z_1,z_2}$ is ensured by Proposition~\ref{univrot}.
Then every $\pi_{z_1,z_2}$ is an irreducible representation, every irreducible representation is unitarily equivalent
to some $\pi_{z_1,z_2},$ and $\pi_{z_{1}^{\prime},z_{2}^{\prime}}$ and $\pi_{z_1,z_2}$ are unitary equivalent if and only if
$(z_{1}^{\prime},z_{2}^{\prime}) \cong_q (z_1,z_2).$
\end{lemma}
{\it Proof.} This is Theorem~$1.9$ in \cite{boca01}. $\qquad \blacksquare$
\begin{lemma}
\label{irredrep2}
Let $\cA$ be a rotation $C^*$--algebra with a frame $(U,V)$ with parameter $\alpha = p/q.$ Let
$\pi \in \textnormal{Hom}(\cA_{\alpha},\cA)$ be the homomorphism that satisfies
$\pi(\widetilde {U}_{\alpha}) = U$ and $\pi(\widetilde {V}_{\alpha}) = V.$
Then the center $\cZ(\cA) = \cC^*(U^q,V^q)$ and $\cA$ is a finitely generated $\cZ(\cA)$--module with basis
$\big\{ \, U^rV^s \, : \, 0 \leq r, s \leq q-1 \, \big\}.$
This means that for every $A \in \cA,$ there exist a unique set of elements $A_{r,s} \in \cZ(\cA), \, 0 \leq r,s \leq q-1$ such that
$$
    A = \sum_{r,s=0}^{q-1} A_{r,s} \, U^r \, V^s.
$$
Moreover, $\pi$ is an isomorphism if and only if $\cZ(\cA)$ is isomorphic to $C(\TT^2).$ This is equivalent,
to the condition that $\Omega\big(\cZ(\cA)\big) = \TT_{c}^{2}.$
\end{lemma}
{\it Proof.} The first two assertions are proved in the discussion following Theorem~$1.10$ in \cite{boca01}.
The third assertion is Proposition~$1.11$ in \cite{boca01}. The last assertion follows from the identifications
explained in the discussion that follows Lemma~\ref{gelfand2}. $\qquad \blacksquare$
\\ \\
Let $\cA$ be a rotation $C^*$--algebra with a frame $(U,V)$ with parameter $\alpha = p/q.$ For $(z_1,z_2) \in \TT_{c}^{2},$
let
$\pi_{z_1,z_2} \in \textnormal{Hom}(\cA_{\alpha},\cM_q)$
denote the homomorphisms defined in Lemma~\ref{irredrep1} and let
$\pi \in \textnormal{Hom}(\cA_{\alpha},\cA)$
denote the homomorphism defined in Lemma~\ref{irredrep2}.
Let
$\widetilde {\pi}$
denote the restriction of $\pi$ to $\cZ(\cA_{\alpha})$ and let
$\widetilde {\pi}_{z_1,z_2}$
denote the restriction of $\pi_{z_1,z_2}$ to $\cZ(\cA_{\alpha}).$
Lemma~\ref{gelfand2} implies that $\cZ(\cA_{\alpha}) = C\big( \Omega(\cZ(\cA_{\alpha})) \big)$
and $\cZ(\cA) = C\big(\Omega(\cZ(\cA))\big).$
Since Lemma~\ref{irredrep2} implies that
$\Omega\big(\cZ(\cA_{\alpha})\big) = \TT_{c}^{2}$
it follows that
$\cZ(\cA_{\alpha}) = C(\TT_{c}^{2}).$
We can use the surjection $\widetilde {\pi}$ to identify $\Omega\big(\cZ(\cA)\big)$ with its image
in $\TT_{c}^{2}$ under the injective map
$\phi \rightarrow \phi \circ \widetilde {\pi}, \ \phi \in \Omega\big(\cZ(\cA)\big),$
and to identify $\widetilde {\pi}$ with the map that
restricts functions in $C(\TT_{c}^{2})$ to the closed subset $\Omega\big(\cZ(\cA)\big).$
Likewise, $\cZ(\cM_q) = \CC,$ $\Omega\big(\cZ(\cM_q)\big)$ consists of a single point,
and $\cZ(\cM_q) = C\big(\Omega(\cZ(\cM_q))\big).$ We may use the surjection
$\widetilde {\pi}_{z_1,z_2}$ to identify $\Omega\big(\cZ(\cM_q)\big)$ with the subset
$\{ \, (z_1,z_2) \, \} \subset \TT_{c}^{2}.$
\begin{pro}
\label{irredrep3}
Let $\cA,$ $\pi$, and $\pi_{z_1,z_2}$ be as above and make also all the identifications as above.
Then there exists a
homomorphism $\rho_{z_1,z_2} \in \textnormal{Hom}(\cA,\cM_q)$
that makes the following diagram commute, i.e. $\pi_{z_1,z_2} = \rho_{z_1,z_2} \circ \pi,$
\begin{displaymath}
\xymatrix{
\cA_{\alpha} \ar[r]^\pi \ar[dr]_{\pi_{z_1,z_2}} & \cA  \ar[d]^{\rho_{z_1,z_2}} \\
  & \cM_q }
\end{displaymath}
if and only if
$(z_1,z_2) \in \Omega\big(\cZ(\cA)\big).$
In this case $\rho_{z_1,z_2}$ is an irreducible representation of $\cA.$ Furthermore,
every irreducible representation of $\cA$ is unitarily equivalent to some $\rho_{z_1,z_2}$ and the
representations $\rho_{z_{1}^{\prime},z_{2}^{\prime}}$ and $\rho_{z_1,z_2}$ are unitary equivalent
if and only if $(z_{1}^{\prime},z_{2}^{\prime}) \cong_q (z_1,z_2).$
\end{pro}
{\it Proof.} For $(z_1,z_2) \in \TT_{c}^{2}$ there exists a $\rho_{z_1,z_2}$ such that
$\pi_{z_1,z_2} = \rho_{z_1,z_2} \circ \pi$
if and only if
\begin{equation}
\label{kernelinclusion}
\textnormal{kernel}(\pi) \subseteq \textnormal{kernel}(\pi_{z_1,z_2}).
\end{equation}
Lemma~\ref{irredrep2} shows that Inclusion~\ref{kernelinclusion} holds if and only if
the following inclusion holds
\begin{equation}
\label{restcenter}
    \textnormal{kernel}(\widetilde {\pi}) \subseteq \textnormal{kernel}(\widetilde {\pi}_{z_1,z_2}).
\end{equation}
Since the map $\widetilde {\pi}$ restricts functions in $C(\TT_{c}^{2})$ to the subset
$\Omega\big(\cZ(\cA)\big)$
and the map $\widetilde {\pi}_{z_1,z_2}$ restricts functions in $C(\TT_{c}^{2})$ to the subset
$\{ \, (z_1,z_2) \, \}$
it follows that Inclusion~\ref{restcenter} holds if and only if $(z_1,z_2) \in \Omega\big(\cZ(\cA)\big).$
The remaining assertions follow from Lemmas \ref{irredrep1}. $\qquad \blacksquare$
\\ \\
We consider the Hilbert space $L^2(\TT^2)$ with the standard orthonormal basis
$$
    \big\{ \xi_{m,n}(x,y) = \xi_{m}(x)\xi_{n}(y) \, : \,  (x,y) \in \TT^2, \, m, n \in \ZZ \, \big\}.
$$
Then the group $SL(2,\ZZ)$ (integer matrices with determinant one) admits the representation
$\Phi \, : \, SL(2,\ZZ) \rightarrow \cB\big(L^2(\TT^2)\big)$ given by
\begin{equation}
\label{unimodrep}
\Phi(g) \, \xi_{m,n} = \xi_{am+bn,cm+dn}, \qquad
g =
\left[ \begin{array}{cc}
a & b \\
c & d
\end{array}\right] \in SL(2,\ZZ), \ \ m, n \in \ZZ.
\end{equation}
This representation is called the Brenken-Watatani automorphic representation of $SL(2,\ZZ)$ on $\cA_{\alpha}$.
\begin{lemma}
\label{conjugation}
$\cA_{\alpha}$ admits a faithful representation on $L^2(\TT^2)$ such that
\begin{eqnarray}
  \widetilde {U}_{\alpha} \, \xi_{m,n} &=& \exp(i\pi \alpha n) \, \xi_{m+1,n}\, , \\
  \widetilde {V}_{\alpha} \, \xi_{m,n} &=& \exp(-i\pi \alpha m) \, \xi_{m,n+1}.
\end{eqnarray}
Furthermore, if $\Phi(g)$ and $g \in SL(2,\ZZ)$ are described by Equation~\ref{unimodrep} then
\begin{eqnarray}
  \Phi(g) \, \widetilde {U}_{\alpha} \, \Phi(g)^{-1} &=& \exp(-i\pi \alpha\, a\, c \, ) \,
    \widetilde {U}_{\alpha}^{a} \, \widetilde {V}_{\alpha}^{c} \, ,\\
  \Phi(g) \, \widetilde {V}_{\alpha} \, \Phi(g)^{-1} &=& \exp(i\pi \alpha\, b\, d\,  ) \,
    \widetilde {U}_{\alpha}^{b} \, \widetilde {V}_{\alpha}^{d}.
\end{eqnarray}
\end{lemma}
{\it Proof.} The commutation relation follows from the computation
\begin{equation}
\label{comm}
    \widetilde {U}_{\alpha} \, \widetilde {V}_{\alpha} \, \xi_{m,n} =
    \exp(-i\pi \alpha m ) \, \widetilde {U}_{\alpha}  \, \xi_{m,n+1} =
    \exp(-i\pi \alpha m ) \, \exp\big[ i \pi \alpha (n+1) \big] \, \xi_{m+1,n+1} =
    \exp(i2\pi \alpha \big )\, \widetilde {V}_{\alpha} \, \widetilde {U}_{\alpha} \, \xi_{m,n}.
\end{equation}
The unitary equivalences follow from almost identical calculations. The first one is
\begin{equation}
\label{equiv}
\Phi(g) \, \widetilde {U}_{\alpha} \, \Phi(g)^{-1} \xi_{m,n} =
\exp \big[ i\pi \alpha (-cm+an)  \big] \, \Phi(g) \, \xi_{dm-bn+1,-cm+an} =
\exp(-i\pi \alpha \, a\, c) \, \widetilde {U}_{\alpha}^{a} \, \widetilde {V}_{\alpha}^{c} \xi_{m,n}.
\end{equation} $\qquad \blacksquare$
\begin{pro}
\label{auto}
Define $\widetilde {W}_{\alpha} = \exp(-i\pi \alpha) \, \widetilde {U}_{\alpha}\, \widetilde {V}_{\alpha}.$
There exists an $L \in \cB\big(L^2(\TT^2)\big)$ such that
\begin{eqnarray}
  L \, \widetilde {U}_{\alpha} \, L^{-1} &=& \widetilde {W}_{\alpha} \, , \\
  L \, \widetilde {V}_{\alpha} \, L^{-1} &=& \widetilde {V}_{\alpha}.
\end{eqnarray}
\end{pro}
{\it Proof.} Follows from Lemma~\ref{conjugation} by choosing $L = \Phi(g)$ where $g$ has entries $b = 0, \, a = c = d = 1.$ $\qquad \blacksquare$
\\ \\
{\bf Remark} Choosing $a = d = 0, \, b = -1, \, c = 1$ gives the Fourier automorphism
$\Phi(g)(\widetilde U_{\alpha}) = \widetilde V_{\alpha}$ and
$\Phi(g)(\widetilde V_{\alpha}) = \widetilde U_{\alpha}^{-1}$
that was used in Corollary~$2.2$ in \cite{boca01} to derive the
Aubry-Andr\'{e} \cite{andreaubry80} identity
$\sigma\big(\widetilde {H}(\alpha,\lambda)\big) = \lambda \, \sigma\big((\widetilde {H}(\alpha,\lambda^{-1})\big)$
where $\widetilde {H}(\alpha,\lambda)$ is defined in Equation~\ref{op2}.
\begin{lemma}
\label{pitheta}
For every $\theta \in [0,1),$ there exists a unique $C^*$--algebra homomorphism
$\pi_{\theta} \, : \, \cA_{\alpha} \rightarrow \cB_{\alpha}$ such that
$\pi_{\theta}(\widetilde {U}_{\alpha}) = \xi_1(\theta)R(\alpha)$
and
$\pi_{\theta}(\widetilde {V}_{\alpha}) = M(\xi_1).$
\end{lemma}
{\it Proof.} Since $\big( \xi_1(\theta)R(\alpha), \, M(\xi_1)\big)$ is a frame with parameter $\alpha$ for $\cB_\alpha,$ the assertion follows from Proposition~\ref{univrot}. $\qquad \blacksquare$
\begin{lemma}
\label{pisimple}
For $\theta \in [0,1)$ the homomorphism $\pi_{\theta}$ is injective if and only if $\alpha$ is irrational.
If $\alpha$ is irrational, then $\pi_{\theta}$ is a bijection, $\cA_{\alpha} \cong \cB_{\alpha},$ and therefore
\begin{equation}
\label{specAirr1}
    \sigma(A) = \sigma\big( \pi_{\theta}(A) \big), \qquad A \in \cA_\alpha, \, \theta \in [0,1).
\end{equation}
\end{lemma}
{\it Proof.} Theorem~$1.10$ in \cite{boca01} implies that if $\alpha$ is irrational then $\cA_{\alpha}$ is simple and hence $\pi_{\theta}$ is injective. If $\alpha$ is rational then $\pi_{\theta}$ is not injective since $\pi_{\theta}\big( \widetilde {U}_{\alpha}^{q} - \xi_1(q \theta) \, I_{\cA_{\alpha}} \big) = 0.$
The second assertion follows immediately. $\qquad \blacksquare$
\begin{pro}
\label{embed}
If $\mu > 0$, $0 \leq \alpha_1 < \alpha_2 < 1,$ and $(U_j,V_j)$ is a frame in $\cA_{\alpha_j}$ for $j = 1, 2,$ then there exists a Hilbert space $\HH$ and injective homomorphisms $\pi_j \, : \, \cA_{\alpha_j} \rightarrow \cB(\HH)$ for $j = 1, 2$ such that
\begin{equation}
\label{ineqalpha12}
    ||\pi_1(U_1) - \pi_2(U_2)|| \ \leq \ 9 \mu \sqrt {2\pi(\alpha_2-\alpha_1)}, \qquad
    ||\pi_1(V_1) - \pi_2(V_2)|| \ \leq \ \frac{27}{\mu} \sqrt{2\pi(\alpha_2-\alpha_1)}.
\end{equation}
\end{pro}
{\it Proof.} Corollary~$3.5$ in \cite{boca01} gives a proof based on the perturbation result (\cite{boca01}, Theorem~$3.1$) which is due to Haagerup and R\o rdam \cite{haagerup95}. $\qquad \blacksquare$
\begin{definition}
\label{azumaya}
A $C^*$--algebra $\cA$ is called an Azumaya $C^*$--algebra if it is isomorphic to the algebra of continuous sections of a matrix
bundle whose base space equals $\Omega\big(\cZ(\cA)\big),$ see \cite{phillips80}. This means that there exists an
integer $q \in \NN$ and a vector bundle with total space $E(\cA)$ and projection map
$\phi \, : \, E(\cA) \rightarrow \Omega\big(\cZ(\cA)\big)$
such that for every
$w \in \Omega\big(\cZ(\cA)\big)$ the fiber $\phi^{-1}(w) \subseteq E(\cA)$
is isomorphic to the algebra $\cM_q.$
Here, each element $A \in \cA$ is identified with a function
$A \, : \, \Omega\big(\cZ(\cA)\big) \rightarrow E(\cA)$
such that $\phi \circ A$
is the identity map on $\Omega\big(\cZ(\cA)\big).$ With this identification,
if $A, B \in \cA$ and $w \in \Omega\big(\cZ(\cA)\big),$ then
$A^*(w) = \big(A(w)\big)^*,$
$(AB)(w) = A(w)B(w),$ and
$(A+B)(w) = A(w) + B(w).$
\end{definition}
\begin{pro}
\label{specrat}
Assume that $q \in \NN$ and that $\cA$ is an Azumaya $C^*$--algebra whose fibers are isomorphic to $\cM_q.$ Then the irreducible
representations of $\cA$ have the form $A \rightarrow A(w)$ for $w \in \Omega\big(\cZ(\cA)\big)$ and where $A$ is identified with a section
of the bundle described in Definition \ref{azumaya}.
If $A \in \cA$ then
\begin{equation}
\label{specAirr2}
    \sigma(A) \ = \, \bigcup_{w \, \in\,  \Omega\left(\cZ(\cA)\right)} \textnormal{eig}\big(A(w)\big).
\end{equation}
If $\Omega\big(\cZ(\cA)\big)$ is connected then for every $A \in \cA,$ $\sigma(A)$ is a union of at most $q$ disjoint connected subsets of $\CC,$
and if $A$ is unitary then $\sigma(A)$ consists of at most $q$ disjoint arcs (or bands) in the circle $\TT_c.$
\end{pro}
{\it Proof.} The bundle representation for rational rotation $C^*$--algebras was constructed in \cite{hoegh81}, elaborated in \cite{bratteli92}, and used to obtain a structure theorem for rational rotation algebras in \cite{stacey98}. Let $I_{q} \in \cM_q$ denote the identity matrix. The second assertion follows
since if $A \in \cA$ is identified with a section of the bundle, then for every $\lambda \in \CC,$ $A-\lambda I_{\cA}$ is invertible if and only if $A(w)-\lambda I_q$ is invertible for every $w \in \Omega\big(\cZ(\cA)\big).$ The second assertion also follows from Proposition~\ref{separate}. The last assertion follows since for every $A \in \cA,$ the set $\textnormal{eig}\left[A(w)\right]$ is a continuous function of $w \in \Omega\big(\cZ(\cA)\big).$ $\qquad \blacksquare$
\\ \\
Azumaya algebras were introduced by Azumaya \cite{azumaya51} (see also \cite{eisenbud00}) and $C^*$--Azumaya algebras are discussed in \cite{phillips80}, \cite{phillips82}. Concrete $C^*$--algebras of operators were used (informally) by Werner Heisenberg and (more formally) by Pascual Jordan in the 1920's to model the algebra of observables in quantum mechanics. A special class of $C^*$--operator algebras, viz. those closed in the weak operator norm topology, were studied by John von Neumann and Francis Murray in a series of papers between 1929--1949. Abstract $C^*$--algebras were invented in 1947 by Irving Segal \cite{segal47} based on the ideas in the seminal 1947 paper by Israel Gelfand and Mark Naimark \cite{gelfandnaimark43}. A $C^*$--algebra $\cA$ is called almost finite (AF) if it contains an increasing sequence $\cA_{n}, \, n \in \NN$ of $C^*$--subalgebras whose union is dense. These algebras were introduced in the seminal paper by Bratteli \cite{bratteli72}. Rotation $C^*$-algebras appear implicitly in Peierls' 1933 paper (\cite{peierls33}, Equations 48--53). Irrational rotation $C^*$--algebras were systematically studied by Rieffel \cite{rieffel81}. In 1980 Pimsner and Voiculescu \cite{pimsner80} showed that for irrational $\alpha$, $\cA_\alpha$ can be embedded in the AF algebra constructed from the continued fraction expansion of $\alpha$ and in 1993 Elliot and Evans \cite{elliotevans93} determined the precise structure of irrational rotation algebras. Applications of rotation algebras (also called noncommutative tori) to the Kronecker Foliation and to the Quantum Hall Effect are discussed by Connes \cite{connes94}.


\end{document}